\documentclass[12pt,a4paper]{article}
\usepackage[utf8]{inputenc}
\usepackage{amsthm}
\usepackage{amsmath}
\usepackage{amsfonts}
\usepackage{amssymb}
\usepackage{gensymb}
\usepackage{graphicx}
\usepackage[dvipsnames]{xcolor}
\usepackage{subcaption}
\usepackage{tikz}
\usepackage{pgfplots}
\usepackage[breaklinks=true]{hyperref}
\usepackage{booktabs}
\usepackage{threeparttable}
\usepackage{array}
\usetikzlibrary{calc}
\usetikzlibrary{positioning,automata,arrows,shapes,patterns}
\usetikzlibrary{decorations.pathreplacing,calligraphy} 
\usepackage{natbib}
\usepackage{dsfont}

\usepackage{geometry}
\newtheorem{theorem}{Theorem}

\newtheorem{definition}{Definition}
\newtheorem{example}{Example}
\newtheorem{lemma}{Lemma}
\newtheorem{proposition}[theorem]{Proposition}

\newtheorem*{proposition*}{Proposition}

\theoremstyle{definition}

\newcommand{\eq}[1]{\begin{equation}#1\end{equation}}
\newcommand{\eqst}[1]{\begin{equation*}#1\end{equation*}}
\newcommand{\al}[1]{\begin{align}#1\end{align}}
\newcommand{\alst}[1]{\begin{align*}#1\end{align*}}
\def\x{\mathbf x}
\def\X{\mathbf X}
\def\p{\mathbf p}

\def\F{\mathcal F}
\def\S{\mathcal S}

\def\B{\mathcal B}

\def\M{\mathcal M}
\def\bnu{\bar{\nu}}

\def\bp{\bar{p}}

\def\hs{\hat{\sigma}}

\DeclareMathOperator*{\argmax}{arg\,max}

\begin{document}
	
\title{\textbf{Dynamic Monopoly Pricing With Multiple Varieties: Trading Up}}

\author{Stefan Buehler, Nicolas Eschenbaum, Severin Lenhard\thanks{Buehler:\ University of St.\thinspace Gallen, Department of Economics, Rosenbergstr.~22, 9000 St.\thinspace Gallen, Switzerland (stefan.buehler@unisg.ch); Eschenbaum:\ Swiss Economics, Ottikerstr.~7, 8006 Zurich, Switzerland (nicolas.eschenbaum@swiss-economics.ch); Lenhard:\ Secretariat of the Swiss Competition Commission and University of St.\thinspace Gallen, Department of Economics, Rosenbergstr.~22, 9000 St.\thinspace Gallen, Switzerland (severin.lenhard@unisg.ch). We thank Maximilian Conze, Alia Gizatulina, Samuel Haefner, Paul Heidhues, Sebastian Kranz, Igor Letina, Ola Mahmoud, Marc Moeller, Georg Noeldeke, Marek Pycia, Armin Schmutzler, Philipp Zahn, and seminar audiences at DICE, the University of Southampton, the University of St.~Gallen, and numerous conferences for helpful discussions and comments. All remaining errors are ours. Stefan Buehler and Nicolas Eschenbaum gratefully acknowledge financial support from the Swiss National Science Foundation through grant No.~100018-178836.} \\\phantom{x}}

\vspace{0.4cm}

\date{\normalsize This draft:\ \today 
}

\maketitle

\begin{abstract}
\noindent This paper studies dynamic monopoly pricing for a broad class of settings that allow for multiple durable, multiple rental, or a mix of varieties. We show that the driving force behind pricing dynamics is the existence of trading-up opportunities. If there are no trading-up opportunities in the static monopoly outcome, then pricing dynamics do not emerge in equilibrium. With trading-up opportunities, pricing dynamics arise until these opportunities are exhausted or the game ends. We characterize the lower bound for the emerging prices and profit and study the conditions under which pricing dynamics end in finite time.
\end{abstract}



\addtocounter{page}{-1}
\thispagestyle{empty}
\newpage
\thispagestyle{plain}

\section{Introduction}

The analysis of dynamic monopoly pricing constitutes an important and long-standing challenge in economic theory. There is a broad consensus that Coasian dynamics are
key for understanding dynamic monopoly pricing: The monopolistic seller of a durable good who cannot commit to future prices has an incentive to lower prices over time, because high-value buyers purchase and leave the market early, whereas low-value buyers remain in the market (negative selection).\footnote{The lack of commitment constrains the monopolist's market power, and in the limit the price converges to marginal cost if all trade takes place in the ``twinkle of an eye'', as conjectured by \cite{Coase1972} and formally established by \cite{Stokey1981}, \cite{Bulow1982}, \cite{Fudenberg1985}, \cite{Gul1986}, and \cite{Ausubel1989}.} As a consequence, forward-looking buyers have an incentive to strategically delay their purchase and wait for lower prices. However, recent research has established that the emergence of Coasian dynamics is not a foregone conclusion.

For example, there is no commitment problem for the seller---and Coasian dynamics do not emerge---if the potential buyers of a durable good have access to an outside option with strictly positive value that ends the game \citep{Board2014}. Similarly, monopoly prices remain constant if high-value rather than low-value consumers remain in the market (positive selection) for a rental good \citep{Tirole2016}.\footnote{If the seller offers a rental good and both negative selection (for non-buyers) and positive selection (for loyal buyers) are at work, then Coasian dynamics for the prices offered to non-buyers lead to ``behavior-based pricing'' \citep{Acquisti2005,Armstrong2006,Fudenberg2007,buehler-eschenbaum}.} Finally, if the seller offers two durable varieties rather than one, then, although Coasian dynamics apply, they do not necessarily lead to marginal cost pricing in the limit \citep{NavaShiraldi2019}. That is, failures of the Coase conjecture have been shown to emerge for different reasons in different settings. 

This paper develops an analytical framework that captures a broad class of monopoly pricing problems, aiming to explain why Coasian dynamics arise in some settings but not in others, and offers a simple approach to determining whether pricing dynamics emerge in Perfect Bayesian Equilibrium (PBE). To fix ideas, consider the following example based on \citet{Hart1988}. A single risk-neutral buyer has a private (per-period) 
 valuation $v^H$ with probability $f$ and $v^L \leq v^H$ with probability $1-f$ for a product. In the static game, a single supplier sets the monopoly price
$$
p^m = \begin{cases}
v^H \text{, if } f v^H > v^L; \\
v^L \text{, if } f v^H \leq v^L.
\end{cases}
$$
Therefore, if $f v^H \leq v^L$, the market clears, and all trading-up opportunities are exhausted. Otherwise, only the high type buys, while the gains from trade with the low type are foregone, leaving trading-up opportunities (Assumption 1 in \citet{Hart1988}). 

In a two-period version of the game where the product is durable and the discount factor is $\delta \in [0,1)$, type $i \in \{H,L\}$ buys in the first period if and only if $(1+\delta) v^i - p^1 \geq \delta (v^i - p^2) \Leftrightarrow v^i \geq p^1- \delta p^2$, where $p^t$ is the price in period $t$. If the seller sets prices to separate buyer types, the profit-maximizing strategy is to set $p^2 = v^L$ and $p^1 = v^H+ \delta v^L$, which results in a profit of $f v^H + \delta v^L$.\footnote{The high type has valuation $(1+\delta) v^H$ and thus obtains an information rent.} If the seller pools both types together, the prices $p_1 = p_2 = (1+\delta)v^L$ are profit-maximizing,\footnote{All prices $p^2\geq v^L$ are profit maximizing and resulting in the same allocation.} resulting in a profit of $(1+\delta) v^L$. Thus, the seller separates types in the repeated game if and only if $f v^H > v^L$. 

This example illustrates that, whenever setting the price to $v^L$ is optimal in the static game, it is optimal to set the same constant price in the repeated version. That is, pricing dynamics do not emerge in the repeated game if there are no trading-up opportunities in the static monopoly outcome. 

Our paper applies this idea to the analysis of a broad class of dynamic monopoly pricing problems that allows for multiple durable, multiple rental, or a mix of varieties. To the best of our knowledge, problems with multiple rental or mixed varieties have not been studied before. The analysis highlights that the driving force behind pricing dynamics---as opposed to the repeated play of static monopoly prices---is the existence of trading-up opportunities: By trading-up the consumer to a higher-valued state, the seller can benefit from splitting up a larger surplus from trade. Our notion of trading up extends the logic of Coasian dynamics, which applies to the prices offered to a non-buyer of a durable good, to the prices offered to a buyer of a non-durable good.

We study a setup with a single seller with constant marginal cost normalized to zero. The seller chooses prices for two varieties of a good, facing a single buyer who is privately informed about her valuations for these varieties. We assume that the seller cannot commit to future prices. In each period, the buyer selects one of three states: she either purchases one of the two varieties or chooses the outside option (no consumption). A fixed set of admissible transitions between these three states governs the choices that are available to the buyer. In particular, if one of the two varieties is not accessible to the buyer throughout the game, our setting reduces to a one-variety problem. If the buyer cannot select the outside option in a given period and must stick with her previous consumption choice, we impose a price of zero for her previous consumption choice to prevent expropriation. It is convenient to think of an absorbing variety as a durable good that can be sold once and for all future periods, whereas a variety that can be purchased in every period separately can be viewed as a rental good. We are interested in characterizing the pricing dynamics in PBE.

We derive three key results. \emph{First}, we show that the seller can do no better than charge static monopoly prices if there are no trading-up opportunities in the static monopoly outcome, regardless of whether durable, rental, or mixed varieties are offered. Intuitively, the result follows because there are no consumer types that can benefit from switching to another consumption path if there are no trading-up opportunities in the static monopoly outcome.  The result implies that Coasian dynamics do not emerge in settings without trading-up opportunities in the static monopoly outcome, irrespective of the seller's commitment ability. The result is reminiscent of \citet{Tirole2016}'s finding that it is optimal to charge the static monopoly price to loyal buyers of a single rental good in a positive selection setting with an absorbing outside option. It is also in line with \citet{Board2014}'s analysis, where trading-up opportunities in the static monopoly outcome are excluded by an additional outside option with a strictly positive valuation for all consumers. \emph{Second}, we show that, after any history at which there are trading-up opportunities, the seller lowers prices until all trading-up opportunities are exhausted. Yet, dynamic prices do not fall below the prices $\bp$ associated with the seller-optimal outcome in the static game that leaves no trading-up opportunities. In addition, the seller's present discounted profit is bounded from below by the repeated static profit $\pi(\bp)$ that leaves no trading-up opportunities, which implies that the seller can obtain a positive profit in various settings.\footnote{In the previous example with one durable product, we have $\bar{p}=v^L$ and profit $\pi(\bar{p})=(1+\delta)v^L.$} We further show that whether or not the pricing dynamics are played out in finite time depends on the setting under study and the lowest values in the support. \emph{Third}, we show that our analytical approach also works for transitional games, where one of the varieties is only indirectly available from the initial state (via the other variety), provided that the associated static games are properly defined.

Our analysis highlights that, for a broad class of dynamic monopoly pricing problems, the pricing dynamics depend on whether the monopoly outcome in the (properly defined) static game leaves trading-up opportunities to the seller. If the monopoly outcome in the static game leaves no trading-up opportunities, then the seller does not face a commitment problem and can implement the repeated static monopoly outcome. Instead, if there are trading-up opportunities in the static monopoly outcome, then the seller lowers prices to trade up the consumer to higher-valued consumption options, and a zero-profit lower bound applies in some settings, but not in general. Our analysis suggests that the essence of Coase's insight generalizes to the pricing of multiple non-durable varieties: pricing dynamics emerge whenever the seller has an incentive to switch the consumer to higher-valued consumption options.

This paper adds to an extensive literature on the pricing of a single durable good \citep[e.g][]{Coase1972, Fudenberg1985, Hart1988, Sobel1991, Kahn1986, Bond1984,Fuchs2010}, of multiple varieties of a durable good \citep[e.g.,][]{NavaShiraldi2019,Board2014}, and of vertically differentiated durable products \citep{Hahn2006,Inderst2008,Takeyama2002}. Our work differs by proposing a unified analytical framework that allows for settings with two rental or mixed varieties, respectively, which have largely gone unnoticed. We further contribute to the analysis of positive selection \citep{Tirole2016} by showing how it can be extended to multiple varieties. In addition, we add to the literature on behavior-based pricing \citep[e.g.][]{Acquisti2005,Armstrong2006,Fudenberg2007,taylor2004,buehler-eschenbaum} by considering multiple varieties. In line with the bulk of previous research, we focus on price posting and abstract from smart contracts \citep{Brzustowski2023} or bundling  \citep{RochetThanassoulis2019}.
Finally, we note that our notion of ``trading up'' is different from ``upselling'' \citep[e.g.,][]{Blattberg2008,Goker2008,Wilkie1998} and offering add-ons \citep{Yu-et-al-2025} because it applies to buyers \textit{and} non-buyers.

The remainder of the paper is organized as follows. Section~\ref{sec-framework} introduces the setup, formalizes the notion of trading-up opportunities, and discusses various nested cases. Section~\ref{sec-analysis} derives a simple skimming property for the unified analytical framework and characterizes dynamic monopoly pricing with and without trading-up opportunities in static monopoly. We illustrate our results with two examples in Section~\ref{sec-example}. Section~\ref{sec-ext-transitional} extends the analysis to transitional games. Section~\ref{sec-conclusion} concludes and offers directions for future research.

\section{Setup}\label{sec-framework}

A monopolist offers two varieties of a good, $a$ and $b$, over discrete time $t=1,...,T$, with  $T\leq\infty$, to a single (risk-neutral) consumer with unit demand for the good in every period. Following \citet{NavaShiraldi2019}, the consumer's value profile $v=(v_a,v_b)$ is fixed, private information, and distributed according to a probability measure $\F$ defined on the unit square $[0,1]^2$. The associated cumulative  distribution is $F$, with density $f$, and $V$ is the support (i.e., the smallest closed set such that $\F([0,1]^2 \setminus V)=0$). 
Let $\mathcal{B}(V)$ be the Borel sigma-algebra of $V$. 
We assume that the measure $\F$ is atomless: for any $E \in \B(V)$ with $\F(E) >0$, there exists $E' \in \B(V)$ with $E' \subset E$ and $0<\F(E')<\F(E)$. Moreover, let the support $V$ be convex. The value of the outside option is normalized to zero, and players share the same discount factor $\delta \in[0,1)$.\footnote{We will work with the natural interpretation of the model that there is a single consumer whose type is unknown. Under the alternative interpretation of a continuum of infinitesimal consumers  indistinguishable to the seller, who observes only the measures of sets of consumers who accept and reject, there is no coordination among infinitesimal consumers  \cite[p.\ 400]{Fudenb-Tirole-1993}. }

In every period $t\geq 1$, the consumer makes a discrete choice $x^t\in X$, where
\alst{X \equiv\{a,b,o\}}
is the set of states, with varieties $a=(1,0)$ and $b=(0,1)$, and the outside option $o=(0,0)$.
Let $x^0\in X$ denote the initial state. A sequence of choices $x^t$ from period~$t$ onward traces out a \emph{consumption path} $\x^t = (x^t,x^{t+1},..., x^T)$ that generates (discounted to period $t$) \emph{total consumption} $\chi(\x^t) = \sum_{\tau=t}^T \delta^{\tau-t} x^\tau$. A consumption path is \emph{admissible} if all transitions along the entire path are within the set of admissible transitions $\Gamma\subset X \times X$, where $\Gamma$ is exogenous and determines how consumers can switch between states from one period to the next. Throughout, we assume that transitions from any given state to itself are always admissible, that is, $(o,o), (a,a), (b,b) \in \Gamma$. In the main part of our analysis, we focus on settings in which each state is either directly accessible from the initial state or not accessible at all. We will consider the extension to settings where one variety is only indirectly accessible via another state (``transitional games'') in Section~\ref{sec-ext-transitional}. A state $x\in X$ is \emph{absorbing} if no other state $x'\neq x$ is accessible
from $x$, that is, if $(x,x')\notin \Gamma$. Let $\X(h^{t})$ be the set of admissible consumption paths after history $h^{t}$. In addition, let $\Delta^t = \sum_{\tau=t}^T \delta^{\tau-t}$ denote the (discounted to period $t$) \emph{number of periods} from $t$ onward. For simplicity, we will omit the time superscript of $\Delta^t$ if $t=1$ from now on.

We call a variety ``durable'' if it is absorbing and can only be sold once and for all future periods. A variety that allows for transitions to other states, in turn, is called a ``rental'' variety. To simplify exposition, we will henceforth refer to the setting with two absorbing varieties and non-absorbing initial state $x^0=o$ as the ``two durables'' setting. Similarly, we will refer to the setting with all transitions being admissible and initial state $x^0 = o$ as the ``two rentals'' setting. Lastly, we will refer to the setting with one rental and one durable variety with initial non-absorbing state $x^0 = o$ as the ``mixed varieties'' setting.

\autoref{fig-non-abs} illustrates two different settings: two rentals (panel (a)), and mixed varieties (panel (b)). The vertices indicate states $X=\{a,b,o\}$, with initial state $x^0=o$, and the arcs and brackets $(x,x')\in \Gamma$ represent admissible transitions. Panel (a) shows that all transitions are admissible in the two rentals setting, that is, $\Gamma = \{(a,a), (b,b) ,(o,o), (a,b), (b,a),\allowbreak (o,a), (o,b),\allowbreak (a,o),(b,o)\}$. Panel (b) indicates that in a  mixed varieties setting where $b$ is the durable variety, the transitions $(b,o), (b,a) \notin \Gamma$ are not admissible because durable varieties are absorbing states.


\begin{figure}[h!]
	\centering
	
	\begin{subfigure}{0.49\textwidth}
    	\centering
    	\caption{Two rentals}
    	\begin{tikzpicture}[thick, node distance = 2.8cm, shorten >= 3pt, shorten <= 3pt, ->]
    	\node[state, initial,inner sep=1pt,minimum size=25pt] (o) {$o$};
    	\node[state, left of=o, below of=o,inner sep=1pt,minimum size=25pt] (a) {$a$};
    	\node[state, right of=o, below of=o,inner sep=1pt,minimum size=25pt] (b) {$b$};
    	\draw (o) edge[bend left=15, right] node{$(o,a)$} (a);
    	\draw (a) edge[bend left=15, left] node{$(a,o)$} (o);
    	\draw (b) edge[bend left=15, below] node{$(b,a)$} (a);
    	\draw (a) edge[bend left=15, below] node{$(a,b)$} (b);
    	\draw (o) edge[bend left=15, right] node{$(o,b)$} (b);
    	\draw (b) edge[bend left=15, left] node{$(b,o)$} (o);
    	\draw (a) edge[loop below] node{$(a,a)$} (a);
    	\draw (b) edge[loop below] node{$(b,b)$} (b);
    	\draw (o) edge[loop above] node{$(o,o)$} (o);
    	\end{tikzpicture}
	\end{subfigure}
    \begin{subfigure}{0.49\textwidth}
    	\centering	
    	\caption{Mixed varieties}
    	\begin{tikzpicture}[thick, node distance = 2.8cm, shorten >= 3pt, shorten <= 3pt, ->]
    	\node[state, initial,inner sep=1pt,minimum size=25pt] (o) {$o$};
    	\node[state, left of=o, below of=o,inner sep=1pt,minimum size=25pt] (a) {$a$};
    	\node[state, right of=o, below of=o,inner sep=1pt,minimum size=25pt] (b) {$b$};
    	\draw (o) edge[bend left=15, right] node{$(o,a)$} (a);
    	\draw (a) edge[bend left=15, left] node{$(a,o)$} (o);
    	\draw (a) edge[below] node{$(a,b)$} (b);
    	\draw (o) edge[right] node{$(o,b)$} (b);
    	\draw (a) edge[loop below] node{$(a,a)$} (a);
    	\draw (b) edge[loop below] node{$(b,b)$} (b);
    	\draw (o) edge[loop above] node{$(o,o)$} (o);
    	\end{tikzpicture}
	\end{subfigure}
 \caption{States and transitions in two different settings}\label{fig-non-abs}
    \end{figure}


\subsection{Prices, Histories, and Solution Concept}
In every period $t$, the monopolist selects a price profile $p^t=(p_a^t,p_b^t) \in [\psi,1]^2$, with $\psi < 0$,\footnote{The assumption on the set of prices ensures that the monopolist's action set is compact.} conditional on seller history $h^t$. The consumer then makes consumption choice $x^t \in X(h^t)$, where $X(h^{t}) \subseteq X$ is the set of states accessible after history $h^t$. A sequence of seller histories is given by $h^1 = \{x^0\}$ and $h^t=\{h^{t-1},p^{t-1},x^{t-1}\}$ for $t\geq 2$, where $h^{t-1}$ is a subhistory of $h^t$. Similarly,  a buyer history is given by $\hat{h}^t=\{h^{t},p^{t}\}$ for $t\geq1$.

The set of period-$t$ seller histories is denoted by $H^t$, and the set of all seller histories by $H=\cup_{t=1}^{T}H^t$. Similarly, the set of period-$t$ buyer histories is denoted by $\hat{H}^t$, and the set of all buyer histories by $\hat{H}=\cup_{t=1}^{T}\hat{H}^t$.

A behavioral strategy for the buyer is denoted by $\hs$ and determines the probability distributions over the consumption choices $x^t\in X$ made at every buyer history $\hat{h}^t$. Formally, $\hs: \hat{H} \times V \to s(X)$, where $s(.)$ denotes the respective simplex. In line with the literature, we assume that, at any possible history, the set of buyer types making the same consumption choice is measurable. A behavioral strategy for the seller is denoted by $\sigma$ and determines the probability distribution over the prices $p^t \in [\psi, 1]^2$ set by the seller after every seller history $h^t$. Formally, $\sigma: H \to s([\psi,1]^2)$. 

A \emph{Perfect Bayesian Equilibrium} (PBE) is a strategy profile $\{\sigma, \hs\}$ and updated beliefs about the buyer's value profile along the various consumption paths, such that actions are optimal given beliefs, and beliefs are derived from actions from Bayes' rule whenever possible.

We partition the consumer type space $V$ depending on the history. Let $V(h^1)=V$ and
$$
V(h^{t+1}) = \left \{v \in V(h^{t}) \Big | x^{t} \in \argmax_{x \in X(h^{t})} (v-p^{t}) \cdot x + \delta U (v,x,\hat{h}^{t}) \right \}, \text{ for } t \geq 1, 
$$
where $x^{t}$ is the last element of seller history $h^{t+1}=\{h^{t},p^{t},x^{t}\}$, $p^{t}$ is the price profile selected by the seller after seller subhistory $h^{t}$, and $U(v,x,\hat{h}^{t})$ is the buyer's continuation value after buyer history $\hat{h}^{t} = \{h^{t},p^{t}\}$. Thus, we obtain $\cup_{h^t \in H^t} V(h^t) = V$ at any time $t$ (i.e., no types are left behind). Accordingly, $\F(V(h^t))$ measures all consumer types with history $h^t$ (i.e., all consumer types choosing $x^{t-1}$, facing price profile $p^{t-1}$ after history $h^{t-1}$). 


Importantly, we assume that if the consumer cannot transition to the outside option from variety $x^{t-1} \in \{a,b\}$ after history $h^t$, then the period-$t$ price for variety $x^{t-1}$ at this history is zero, $p^t_{x^{t-1}}(h^t) = 0$. This assumption is consistent with the natural interpretation of an absorbing variety as a durable good and excludes the expropriation of a ``captured'' buyer. Let $\rho(\p^t, \x^t) = \sum_{\tau=t}^T \delta^{\tau} (p^\tau \cdot x^\tau)$ denote the (discounted to period $t$) \emph{total payment} made along consumption path $\x^t$ with price path $\p^t=(p^t,p^{t+1},...,p^T)$ after history $h^t$. Similarly, let $\nu(v,\x^t) = v \cdot \chi(\x^t)$ be the (discounted to period $t$) \emph{total value} obtained by a buyer with value profile $v$ along consumption path $\x^t$ after history $h^t$. We can then write the (present discounted) \emph{net value} obtained by a buyer with value profile $v$ along consumption path $\x=\x^1$ and the path of price profiles $\p=\p^1$ compactly as $\nu(v,\x) - \rho(\p ,\x)$.


\subsection{Trading-up opportunities}

We introduce the following definition.

\begin{definition}[\textbf{Trading-up opportunity}]\label{definition-tuo}
The seller has a trading-up opportunity at history $h^t=\{h^{t-1}, p^{t-1}, x^{t-1} \} $ if there is a positive measure of consumer types who can transition to a strictly higher-valued state,
$$
\F\left( \Big \{ v \in V(h^t) ~\Big| x^{t-1} \notin \argmax_{x \in X(h^t)} v\cdot x \Big \} \right) > 0\text{.}
$$
\end{definition}

Definition~\ref{definition-tuo} formalizes the notion that, for a trading-up opportunity to exist for the seller at history $h^t$, a higher-valued consumption option than $x^{t-1}$ must be accessible to the buyer.

For later reference, we let $\Omega$ denote the set of price profiles $p=(p_a,p_b)$ that induce an allocation which leaves no trading-up opportunities for the seller in the static game,
\begin{equation*}
  \Omega = \left\{p \in [\psi,1]^2 \Big| \F\left( \Big \{ v \in V (\{x^0, p, x^1 \}) ~\Big| x^{1} \notin \argmax_{x \in X(h^2)} v\cdot x \Big \} \right) =0 \right\}.
\end{equation*}
Intuitively, any price profile $p\in \Omega$ must induce an allocation in the static game where the consumer selects either an absorbing state or the most-preferred state among those that are accessible from the initial state. Thus, in a setting with two durables, any price profile $p \in \Omega$ must implement market-clearing. In a setting with two rentals, in turn, any price profile $p \in \Omega$ must implement market-clearing and efficiency.\footnote{We follow \cite{NavaShiraldi2019} in referring to price profiles which ensure that all buyers choose their preferred (accessible) variety as \emph{efficient} because they maximize total welfare when marginal costs are normalized to zero.} 

Because $\F$ is atomless, the static demand for variety $i \neq j \in \{a,b\}$ satisfies $d_i(p) = \F(v \in V | v_i-p_i \geq \max\{v_j - p_j, 0 \})$, resulting in the static profit $\pi(p) = d_a(p) p_a + d_b(p) p_b$. We let $p^m \in \argmax \pi(p)$ denote the static profit-maximizing price profile.

Finally, we let $\bar{p}\in \Omega$ denote a price profile that is associated with the maximum of the profit obtainable in the static game conditional on leaving no trading-up opportunities, $\pi(\bar{p}) = \max \pi(p)$ s.t.\ $p \in \Omega$. This profit maximum exists provided that $\Omega \neq \emptyset$, which is guaranteed since $p = (0,0) \in \Omega$ (we consider the extension to transitional games in \autoref{sec-ext-transitional}).

\begin{figure}[!ht]
\caption{Demand segments in the static game for given $p$ (panel (a)), and profiles $p\in \Omega$ with full support (panel (b)) or linear support (panel (c)) for two rentals}
\label{fig-tuo-threedist}
\centering
   \begin{subfigure}{0.3\textwidth}
    \centering
    \caption{Static demand}
        \begin{tikzpicture} [scale=0.3]
        \draw[line width=0.6mm,->] (-0.5,0) -- (11.5,0);
        \draw[line width=0.6mm,->] (0,-0.5) -- (0,11.5);
        \draw[line width=0.4mm] (10,0) -- (10,10) -- (0,10);
        \draw[fill, black] (5,5) circle(0.2cm);
        \node at (5.7,4.7) {$p$};
        \node at (-0.75,5) {$p_b$};
        \node at (5,-0.75) {$p_a$};
        \draw[thick] (5,0) -- (5,5) -- (0,5);
        \draw[thick] (5,5) -- (10,10);
        \fill[white,opacity=0.2] (0,0) -- (5,0,0) -- (5,5) -- (0,5);
        \fill[CornflowerBlue,opacity=0.2] (5,0) -- (5,5) -- (10,10) -- (10,0);
        \fill[ForestGreen,opacity=0.2] (0,5) -- (5,5) -- (10,10) -- (0,10);
        \node at (2.5,2.5) {$x=o$};
        \node at (4,7.75) {$x=b$};
        \node at (7.5,2.5) {$x=a$};
        \node at (-0.5,10) {$1$};
        \node at (10,-0.7) {$1$};
        \node at (11,-1) {$v_a$};
        \node at (-1.1,11) {$v_b$};
        \end{tikzpicture}
    \label{fig-segments}
    \end{subfigure}
    \hfill
    \begin{subfigure}{0.3\textwidth}
    \centering
    \caption{$p\in \Omega$, full support}
        \begin{tikzpicture} [scale=0.3]
        \draw[line width=0.6mm,->] (-0.5,0) -- (11.5,0);
        \draw[line width=0.6mm,->] (0,-0.5) -- (0,11.5);
        \draw[line width=0.4mm] (10,0) -- (10,10) -- (0,10);
        \fill[blue,opacity=0.1] (0,0) -- (10,0) -- (10,10) -- (0,10);
        \node[text=blue] at (5,5) {$V$};
        \draw[line width = 0.5mm, red] (0,0) -- (-2,-2);
        \node at (-0.5,10) {$1$};
        \node at (10,-0.7) {$1$};
        \node at (11,-1) {$v_a$};
        \node at (-1.1,11) {$v_b$};
        \end{tikzpicture}
    \label{fig-tuo-full}
    \end{subfigure}
    \hfill
    \begin{subfigure}{0.3\textwidth}
    \centering
    \caption{$p\in \Omega$, linear support}
        \begin{tikzpicture} [scale=0.3]
        \draw[line width=0.6mm,->] (-0.5,0) -- (11.5,0);
        \draw[line width=0.6mm,->] (0,-0.5) -- (0,11.5);
        \draw[line width=0.4mm] (10,0) -- (10,10) -- (0,10);
        \draw[line width = 0.5mm, blue] (2,0) -- (10,8);
        \node[text=blue] at (5.5,4.5) {$V$};
        \fill[red,opacity=0.4] (-2,-2) -- (-2,10) -- (2,10) -- (2,0) -- (0,-2);
        \node at (-0.5,10) {$1$};
        \node at (10,-0.7) {$1$};
        \node at (11,-1) {$v_a$};
        \node at (-1.1,11) {$v_b$};
        \end{tikzpicture}
    \label{fig-tuo-vertical}
    \end{subfigure}
\end{figure}

\autoref{fig-tuo-threedist} illustrates for the two rentals setting how buyer types self-select in the static game for a given price profile $p$, and depicts price profiles (in red) that satisfy $p\in \Omega$ for two different supports. Specifically, panel (a) shows the static demand segments for the price profile $p=(0.5,0.5)$ and indicates, for instance, that all consumer types with values $v_i < p_i, \forall i$ choose the outside option, $x = o$. Panels (b) and (c) depict price profiles that leave no trading-up opportunities with full and linear support, respectively. For two rentals, $p \in \Omega$ requires that all consumer types choose their most-preferred variety, as otherwise there are trading-up opportunities from one variety to the other, or from the initial state to each variety. Thus, with full support only non-positive price profiles on the diagonal satisfy $p \in \Omega$ (panel (b)), whereas with an increasing linear support that lies to the right of the diagonal through the type space (panel (c)), any price profile that ensures $x=a$ for all types in the support satisfies $p \in \Omega$, since all buyer types prefer $a$ to $b$ (``vertical differentiation'').

\subsection{Nested cases}\label{subsec-applications}

Our setup covers a broad class of dynamic monopoly pricing settings that can be characterized by the tuple $(x^0,\Gamma, \F)$.
It is convenient to illustrate these settings in two complementary graphs: one showing the accessible states and admissible transitions, and one showing the support $V$ of the value profiles and the price profiles $p\in\Omega$ that leave no trading-up opportunities, respectively. \autoref{fig-applications} provides three examples that are drawn for a full support $V$ on the unit square $[0,1]^2$.

In a setting with a single durable variety $a$ and full support (\autoref{fig-one-durable}), $p \in \Omega$ requires that $p_a \leq 0$ (whereas $p_b$ remains unrestricted), which implies that $\pi(\bp) =0$. In a positive selection setting with a single variety $a$ and initial state $x^0=a$ (\autoref{fig-tirole}), in turn, all price profiles satisfy $p \in \Omega$, which implies that $\pi(\bp) = \pi(p^m)>0$ with full support. Finally, in a mixed setting with rental variety $a$, durable variety $b$, and initial state $x^0=o$ (\autoref{fig-mix}), $p \in \Omega$ requires that the price of the durable variety $b$ is non-positive, whereas the price of the rental variety $a$ can be positive and thus $\pi(\bp) > 0$ with full support.

\begin{figure}[h!]
	\centering
	\caption{Three examples: Accessible states and admissible transitions (left), and price profiles $p\in \Omega$ with a full support (right)}\label{fig-applications}
	\vspace{1em}
	\begin{subfigure}{\textwidth}
	\caption{Single durable variety $a$, with initial state $x^0=o$}
	\vspace{-0.5em}
	\centering
    	\begin{subfigure}{0.53\textwidth}
    	\centering
    	\begin{tikzpicture}[thick, node distance = 3cm, shorten >= 3pt, shorten <= 3pt, ->]
    	\node[state, initial,inner sep=1pt,minimum size=25pt] (o) {$o$};
    	\node[state, right of=o,inner sep=1pt,minimum size=25pt] (a) {$a$};
    	\draw (o) edge node[above]{$(o,a)$} (a);
    	\draw (o) edge[loop above] node{$(o,o)$} (o);
    	\draw (a) edge[loop above] node{$(a,a)$} (a);
    	\end{tikzpicture}
    	\end{subfigure}
	\hfill
    	\begin{subfigure}{0.4\textwidth}
        \centering
        \begin{tikzpicture} [scale=0.25]
        \fill[red,opacity=0.4] (-2,10) -- (0,10) -- (0,-2) -- (-2,-2);
        \draw[line width=0.6mm,->] (-1,0) -- (11.75,0);
        \draw[line width=0.6mm,->] (0,-1) -- (0,11.75);
        \draw[line width=0.4mm] (10,0) -- (10,10) -- (0,10);
        \fill[blue,opacity=0.1] (0,0) -- (10,0) -- (10,10) -- (0,10);
        \node[text=blue] at (5,5) {$V$};
        \node at (-0.75,10) {$1$};
        \node at (10,-0.9) {$1$};
        \node at (11.25,-1.25) {$v_a$};
        \node at (-1.25,11.25) {$v_b$};
        \end{tikzpicture}
        \end{subfigure}
    \label{fig-one-durable}
    \end{subfigure}
 	\vspace{2em}

	\begin{subfigure}{\textwidth}
	\caption{Positive selection, with single variety $a$ and initial state $x^0=a$}
	\vspace{-0.5em}	
	\centering
    	\begin{subfigure}{0.53\textwidth}
    	\centering
    	\begin{tikzpicture}[thick, node distance = 3cm, shorten >= 3pt, shorten <= 3pt, ->]
    	\node[state, initial,inner sep=1pt,minimum size=25pt] (a) {$a$};
    	\node[state, right of=a,inner sep=1pt,minimum size=25pt] (o) {$o$};
    	\draw (a) edge[above] node{$(a,o)$} (o);
    	\draw (a) edge[loop above] node{$(a,a)$} (a);
    	\draw (o) edge[loop above] node{$(o,o)$} (o);
    	\end{tikzpicture}
    	\end{subfigure}	
    \hfill
    	\begin{subfigure}{0.4\textwidth}
        \centering
        \begin{tikzpicture} [scale=0.25]
        \fill[red,opacity=0.4] (-2,10) -- (10,10) -- (10,-2) -- (-2,-2);
        \draw[line width=0.6mm,->] (-1,0) -- (11.75,0);
        \draw[line width=0.6mm,->] (0,-1) -- (0,11.75);
        \draw[line width=0.4mm] (10,0) -- (10,10) -- (0,10);
        \fill[blue,opacity=0.1] (0,0) -- (10,0) -- (10,10) -- (0,10);
        \node[text=blue] at (5,5) {$V$};
        \node at (-0.75,10) {$1$};
        \node at (10,-0.9) {$1$};
        \node at (11.25,-1.25) {$v_a$};
        \node at (-1.25,11.25) {$v_b$};
        \end{tikzpicture}
        \end{subfigure}
	\label{fig-tirole}	
	\end{subfigure}
	\vspace{2em}

    \begin{subfigure}{\textwidth}
    \caption{Mixed setting, with rental variety $a$, durable variety $b$, and initial state $x^0 = o$}
	\vspace{-0.5em}
    	\begin{subfigure}{0.49\textwidth}
    	\centering	
    	\begin{tikzpicture}[thick, node distance = 2.5cm, shorten >= 3pt, shorten <= 3pt, ->]
    	\node[state, initial,inner sep=1pt,minimum size=25pt] (o) {$o$};
    	\node[state, left of=o, below of=o,inner sep=1pt,minimum size=25pt] (a) {$a$};
    	\node[state, right of=o, below of=o,inner sep=1pt,minimum size=25pt] (b) {$b$};
    	\draw (o) edge[bend left=15, right] node{$(o,a)$} (a);
    	\draw (a) edge[bend left=15, left] node{$(a,o)$} (o);
    	\draw (a) edge[below] node{$(a,b)$} (b);
    	\draw (o) edge[right] node{$(o,b)$} (b);
    	\draw (a) edge[loop left] node{$(a,a)$} (a);
    	\draw (b) edge[loop right] node{$(b,b)$} (b);
    	\draw (o) edge[loop above] node{$(o,o)$} (o);
    	\end{tikzpicture}
    	\end{subfigure}
	\hfill
    	\begin{subfigure}{0.4\textwidth}
        \centering
        \begin{tikzpicture} [scale=0.25]
        \fill[red,opacity=0.4] (-2,-2) -- (0,0) -- (10,0) -- (10,-2);
        \draw[line width=0.6mm,->] (-1,0) -- (11.75,0);
        \draw[line width=0.6mm,->] (0,-1) -- (0,11.75);
        \draw[line width=0.4mm] (10,0) -- (10,10) -- (0,10);
        \fill[blue,opacity=0.1] (0,0) -- (10,0) -- (10,10) -- (0,10);
        \node[text=blue] at (5,5) {$V$};
        \node at (-0.75,10) {$1$};
        \node at (10,-0.9) {$1$};
        \node at (11.25,-1.25) {$v_a$};
        \node at (-1.25,11.25) {$v_b$};
        \end{tikzpicture}
        \end{subfigure}
	\label{fig-mix}
	\end{subfigure}
\end{figure}

\clearpage

\section{Analysis}\label{sec-analysis}

We now characterize dynamic monopoly pricing within the framework introduced above. We proceed in three steps. First, we provide a simple skimming result. Second, we introduce a convenient way of representing the seller's profit. Finally, we analyze optimal pricing with and without trading-up opportunities.

\subsection{Skimming}\label{sec-skimming}

We first show that the value profiles of consumer types who make the same consumption choice satisfy an intuitive sorting condition (all proofs are relegated to the Appendix).

\begin{lemma}[\textbf{Skimming}]\label{lemma-skimming}
Consider consumer types with a common history $h^t$. If a consumer type with value profile $v$ obtains a higher net value along path $\x_k^t$ than along path $\x_l^t$, with total consumption $\chi(\x_k^t) \neq \chi(\x_l^t)$, then so does another consumer type with value profile $\tilde{v} \neq v$ such that
\eq{\label{equ-skim}
  (\tilde{v}-v) \cdot (\chi(\x_k^t)-\chi(\x_l^t)) \geq 0 \text{.}
}
\end{lemma}

The result states the skimming condition in terms of the (discounted to period $t$) total consumption levels obtained along two different consumption paths.\footnote{An analogous result can be stated in terms of the consumption choices at time $t$ for any history $h^t$ (rather than the total consumption levels obtained along two consumption paths).} This condition shows that for two consumer types to have the same preferences over the net values generated along two different consumption paths, the value profiles and the total consumption levels along the two paths must be aligned. That is, it is generally not sufficient for a type to have strictly higher values for both varieties to satisfy the condition; instead, the relative values $(v_a-v_b)$ must be considered. For example, if following path $\x_k^t$ instead of $\x_l^t$ implies obtaining relatively less consumption of $a$ and relatively more consumption of $b$, and type $v$ chooses path $\x_k^t$ over $\x_l^t$, then only types $\tilde{v}$ who do not prefer $a$ relatively more than $b$ compared to type $v$ will make the same choice. However, if path $\x_k^t$ implies obtaining more consumption of $a$ compared to path $\x_l^t$, while the consumption of $b$ is equal along the two paths, then $\tilde{v}_a > v_a$ is sufficient for type $\tilde{v}$ to have the same preference. Hence, restricting the set of admissible consumption paths makes it easier to satisfy the skimming condition. 

Figure~\ref{fig-lemma1} illustrates the skimming condition~(\ref{equ-skim}), which characterizes a dot product and restricts the angle between the two direction vectors $(\tilde{v}-v)$ and $\left(\chi(\x_k^t)-\chi(\x_l^t)\right)$ to a maximum of 90\degree. For the particular difference $\left(\chi(\x_k^t)-\chi(\x_l^t)\right)$ depicted, only value profiles in the shaded area (including $(v,\tilde{v}$)) satisfy the condition. 

\begin{figure}[ht]
	\centering
	\caption{Illustration of the skimming condition (\ref{equ-skim})}	\label{fig-lemma1}
	\begin{tikzpicture} [scale=0.5]
        \draw[line width=0.6mm,->] (-0.5,0) -- (11.5,0);
        \draw[line width=0.6mm,->] (0,-0.5) -- (0,11.5);
        \draw[line width=0.4mm] (10,0) -- (10,10) -- (0,10);
        \node at (-0.5,10) {$1$};
        \node at (10,-0.7) {$1$};
        \node at (11,-1) {$v_a$};
        \node at (-1.1,11) {$v_b$};
        \draw[line width=0.4mm,black,-stealth](0,0)--(-5,10) node[anchor=south]{$\chi(\x_k^t)-\chi(\x_l^t)$};
        \draw[line width=0.4mm, dotted] (-2,4) -- (10,10);
        \filldraw (4,7) circle[radius=1mm] node[anchor=north]{$v$};
        \filldraw (1,8) circle[radius=1mm] node[anchor=north]{$\tilde{v}$};
        \draw[line width=0.4mm, black,-stealth] (4,7)--(1,8);
        \fill[black,opacity=0.1] (0,5) -- (10,10) -- (0,10) -- (0,5);
        \draw[line width=0.2mm] (-2.25,4.5) -- (-1.75,4.75) -- (-1.5,4.25);
        \filldraw (-1.875,4.375) circle[radius=0.2mm];
	\end{tikzpicture}
\end{figure}

\subsection{Profit}

Any strategy profile $\{\sigma, \hs\}$ gives rise to sequences of prices and consumption choices that can be computed recursively. We can, thus, measure consumer types with the same history $h^t$, following the same consumption path $\x^t$, as 
$$
\S\left(\x^t | h^t, \sigma, \hs \right) = \F \left( \{ v \in V(h^t) | \text{$(\sigma, \hs)$ imply that $v$ follows $\x^t$} \} \right)
$$
Accordingly, the seller cannot distinguish between these types. 

The seller's present discounted (future) profit at history $h^t$ can be expressed in terms of the strategy-contingent payments made along the admissible consumption paths $\x^t\in \X(h^t)$ by the appropriate measures of consumer types,

\eq{\label{seller-profit}
 \Pi(\sigma,\hs | h^t) =\sum_{\x^t\in \X(h^t)}\rho(\p^t, \x^t|\sigma,\hs, h^t) \S \left( \x^t | \sigma,\hs,h^t\right) \text{,}
}
where $\rho(\p^t, \x^t | \sigma,\hs,h^t)$ is the total strategy-contingent payment along consumption path $\x^t$ after history $h^t$, and $\S(\x^t |\sigma,\hs, h^t)$ is the strategy-contingent measure of types that follow consumption path $\x^t$ after history $h^t$.

\subsection{Pricing without trading-up opportunities}\label{sec-no-TUO}

Borrowing terminology from \cite{Board2014}, we say that the seller and the buyer follow \emph{monopoly strategies} $\{\sigma^m,\hat{\sigma}^m\}$ if, in every period $t$, the seller plays static monopoly prices $p^m$ and the buyer behaves as if she were myopic. That is, in every period $t$,
\begin{itemize}
    \item[(i)] the seller charges $p_i^m$ for rental variety $i\in\{a,b\}$, and the buyer purchases rental variety $i$ if $v_i-p_i^m\geq \max\{v_j-p_j^m,0\}$, $j\neq i$;
    \item[(ii)] the seller charges $p_i^m\Delta ^t$ for durable variety $i\in\{a,b\}$ if the buyer has not yet purchased variety $i$ (and zero otherwise), and the buyer purchases durable variety $i$ if $(v_i-p_i^m)\Delta^t\geq \max\{ (v_j-p_j^m)\Delta^t,0\}$, $j\neq i$.
    \end{itemize}

We can then state the following result.

\begin{proposition}[\textbf{Repeated static monopoly}]\label{prop-1}
Suppose there are no trading-up opportunities in the static monopoly outcome, that is, $p^m\in\Omega$.
Then, in any PBE,
\begin{itemize}
    \item[(i)] the seller can do no better than follow the monopoly strategy along the equilibrium path.
    \item[(ii)] the seller's profit equals the (present discounted) sum of the repeated static monopoly profit, that is, $\Pi = \pi(p^m)\Delta$.
\end{itemize}
\end{proposition}

Proposition~\ref{prop-1} shows that the seller can do no better than follow a monopoly strategy if there are no trading-up opportunities in the static monopoly outcome, regardless of whether durable, rental, or mixed varieties are offered. Intuitively, the result follows because the absence of trading-up opportunities in the static monopoly outcome implies that the seller cannot benefit from inducing the buyer to switch to an alternative consumption path: the resulting short-term loss relative to the static monopoly profit cannot be recouped in the future, because future prices cannot lead to a higher profit than the static monopoly prices that leave no trading-up opportunities.

Importantly, the result implies that Coasian dynamics do \textit{not} emerge in settings without trading-up opportunities in the static monopoly outcome, irrespective of the seller's commitment ability. Hence, it suffices to work out the static monopoly outcome to determine the outcome of the dynamic game for these settings. To be sure, \cite{Tirole2016} has already established this result for a positive-selection setting with a single rental variety and an absorbing outside option with value zero, where trading-up opportunities in static monopoly are excluded by construction.\footnote{See \autoref{fig-tirole} for an illustration of positive selection with a single variety $a$.} Proposition~\ref{prop-1} shows that the result generalizes to settings with multiple varieties under appropriate assumptions. For instance, the result trivially emerges if the initial state is absorbing, regardless of the varieties offered by the seller. More interestingly, if the initial state is non-absorbing and the outside option is absorbing, then the result emerges only if the initial state (i.e., a rental variety) is also the most-preferred state of all consumer types in the support (otherwise, there would be trading-up opportunities). That is, in settings with mixed varieties or two rental varieties, the rental variety in the initial state must be the most-preferred variety for all consumer types in the support.

\cite{Board2014} study a related setting with a single durable variety and a non-absorbing outside option with value zero as the initial state. Clearly, Proposition~\ref{prop-1} should not be expected to hold in this setting, because there are trading-up opportunities in the static monopoly outcome. Nevertheless, these authors show that Coasian dynamics do not emerge if all consumer types have costless access to \textit{another} absorbing outside option with strictly positive value. Their result is consistent with Proposition~\ref{prop-1} because the additional outside option effectively eliminates trading-up opportunities in the static monopoly outcome. The reason is that all types that do not buy at static monopoly prices prefer the costless additional outside option with strictly positive value to the initial state with value zero, leaving no types in the initial state, and hence no trading-up opportunities.

\subsection{Pricing with trading-up opportunities}\label{sec-TUO}

We now consider settings with trading-up opportunities in the static monopoly outcome. The classic example is the case of a single durable good with non-absorbing initial state  $x^0=o$, but similar trading-up opportunities emerge with multiple durable, rental, or mixed varieties. 

In the case of a single durable good, it is well-known that the seller obtains a strictly positive profit if the lowest valuation is above marginal cost---the so-called ``gap case'' \citep[pp.\ 408]{Fudenb-Tirole-1993}. We first show that this result carries over to settings with multiple durable, rental, or mixed varieties: the seller can obtain a strictly positive profit for certain measures $\F$ of the consumer's value profile by trading up all types at once.

\begin{lemma}
If the minimal value of at least one variety is strictly positive, then $\pi(\bar{p})>0$.  
\label{cor1}
\end{lemma}

That is, regardless of whether durable, rental, or mixed varieties are offered, the seller obtains a strictly positive profit if there is a gap for at least one of the varieties. The result will be useful for characterizing when pricing dynamics come to an end.

Next, we show that the seller engages in dynamic pricing after any history at which trading-up opportunities exist until all trading-up opportunities are exhausted. We characterize these pricing dynamics and provide conditions under which they are played out in finite time.

\begin{proposition}[\textbf{Pricing dynamics}]\label{prop-3}
Suppose there are trading-up opportunities in the static monopoly outcome. Then, in any PBE,
\begin{enumerate}
    \item[(i)] the seller trades up a positive measure of types along the equilibrium path after any history $h^t$ with trading-up opportunities.
    \item[(ii)] the seller will never set prices below $\bar{p}$ after any history $h^t$, implying that the seller's (present discounted) profit satisfies $\Pi \geq \pi(\bp) \Delta$.
    \item[(iii)] all trading-up opportunities are exhausted in finite time $t\leq T$ if the minimal value of at least one variety is strictly positive, and $T$ is sufficiently large.
\end{enumerate}
\end{proposition}

Proposition \ref{prop-3} demonstrates that the driving force behind pricing dynamics is the existence of trading-up opportunities. For a seller who faces trading-up opportunities and lacks commitment, it is profitable to trade up positive masses of types to higher-valued consumption options, thereby extracting a larger surplus. Consequently, the seller changes prices to trade up more and more types as the game progresses. Since no price profile $p \in \Omega$ leaves any trading-up opportunities, neither in the static nor in the dynamic game, the seller does not sell at prices below $\bar{p}$. Hence, the dynamics come to an end at prices $\bar{p}$, provided that transitions to the respective consumption options are admissible.\footnote{Note that any price is optimal for a non-admissible consumption option.} This implies that the seller's profit in the absence of commitment is bounded from below at $\pi(\bar{p})\Delta$, which may be strictly positive. The time it takes for the price dynamics to play out depends on the setting under study. However, for all trading-up opportunities to be exhausted in finite time, the minimum value for at least one variety must be strictly positive, and the number of periods of play must be sufficiently large.

To understand the intuition for statement (iii), observe that since it is optimal for the seller to engage in trading-up at any history with trading-up opportunities, the seller must decide whether to trade up some or \emph{all} types. The more types the seller has already traded up in previous periods, the smaller is the extra surplus that can be extracted from the remaining types who can still be traded up. Eventually, it no longer pays for the seller to delay the trading up of some lower-value types in order to trade up higher-valued types earlier on at higher prices, and the seller trades up all remaining consumers instantaneously. But for this to occur in finite time, the seller must be able to strictly increase profit by trading up all types at once. That is, if the minimum value of at least one variety is strictly positive and there are sufficiently many periods of play, the seller will trade up all types in finite time. Otherwise---in the ``no gaps case''---the pricing dynamics may continue indefinitely.

\section{Examples}\label{sec-example}
We present two examples with linear support to illustrate Propositions \ref{prop-1} and \ref{prop-3}. Specifically, we consider $T=2$ periods and let the initial state be the non-absorbing outside option, $x^0 = o$. For each example, we examine three different settings: (i) Two durables, (ii) two rentals; and (iii) mixed varieties. 

    \begin{figure}[h!]
    \caption{Two examples with linear support}
    \begin{subfigure}{0.49\textwidth}
    \caption{Repeated static monopoly}
    \centering
        \begin{tikzpicture} [scale=0.5]
        \draw[line width=0.6mm,->] (-0.5,0) -- (11.5,0);
        \draw[line width=0.6mm,->] (0,-0.5) -- (0,11.5);
        \draw[line width=0.3mm] (10,0) -- (10,10) -- (0,10);
        \draw[line width = 0.6mm, blue] (10, 0) -- (0,10);
        \draw[line width = 0.3 mm, dotted] (0, 5) -- (5,5);
        \draw[line width = 0.3 mm, dotted] (5, 0) -- (5,5);
        \node[text=blue] at (3,8) {$V$};
        \node at (-0.5,10) {$1$};
        \node at (10,-0.7) {$1$};
        \node at (11,-1) {$v_a$};
        \node at (-1.1,11) {$v_b$};
        \node at (5,-1) {$\frac{1}{2}$};
        \node at (-1,5) {$\frac{1}{2}$};
        \end{tikzpicture}
    \end{subfigure}%
    \begin{subfigure}{0.49 \textwidth}
    \caption{Pricing dynamics.}
    \centering
        \begin{tikzpicture} [scale=0.5]
        \draw[line width=0.6mm,->] (-0.5,0) -- (11.5,0);
        \draw[line width=0.6mm,->] (0,-0.5) -- (0,11.5);
        \draw[line width=0.3mm] (10,0) -- (10,10) -- (0,10);
        \draw[line width = 0.6mm, blue] (0, 0) -- (10,10);
        \draw[line width = 0.3 mm, dotted] (0, 3.57) -- (3.57,3.57);
        \draw[line width = 0.3 mm, dotted] (3.57, 0) -- (3.57,3.57);
        \draw[line width = 0.3 mm, dotted] (0, 7.14) -- (7.14,7.14);
        \draw[line width = 0.3 mm, dotted] (7.14, 0) -- (7.14,7.14);
        \draw[line width = 0.2 mm, decorate, decoration={brace,mirror,raise=1 pt}, text=blue]  (7.14,7.14) -- node[below right] {$V_1$}  (10,10);
        \draw[line width = 0.2 mm, decorate, decoration={brace,mirror,raise=1 pt}, text=blue]  (3.57,3.57) -- node[below right] {$V_2$}  (7.14,7.14);
        \node at (-0.5,10) {$1$};
        \node at (10,-0.7) {$1$};
        \node at (11,-1) {$v_a$};
        \node at (-1.1,11) {$v_b$};
        \node at (7.14,-1) {$w$};
        \node at (3.57,-1) {$p^2$};
        \node at (-1,7.14) {$w$};
        \node at (-1,3.57) {$p^2$};
        \end{tikzpicture}
    \end{subfigure}
    \label{fig-example}
    \end{figure}

For the first example, assume that consumer types are uniformly distributed on $V=\{v \in [0,1]^2 | v_a + v_b = 1 \}$, that is, $\F(E) = \int_E 1/\sqrt{2} d \mu(v)$, for $E \in \mathcal{B}(V)$, where $\mu$ is the Lebesgue measure. We derive the following result in Appendix \ref{sec-online-appendix}, which is illustrated in the left panel of Figure~\ref{fig-example}.

\begin{example}[\textbf{Repeated static monopoly}; $T=2$]\label{example-a}
 Let $x^0=o$ and assume that consumer types are uniformly distributed on $V=\{v \in [0,1]^2  | v_a + v_b = 1 \}$. Then, for two durables, two rentals, or mixed varieties, there exists a PBE where the repeated static monopoly outcome emerges and all consumer types are traded-up in $t=1$.
\end{example}

Regardless of the setting, by setting prices such that $p^1_a + p^1_b\geq1$, the supplier can split the set of consumer types into two separate segments that give rise to independent demands (i.e., the demand for variety  $i$ is not affected by the price of variety $j\neq i$). In both segments, the consumer's valuation is uniformly distributed on $[\frac{1}{2},1]$, and the static monopoly price profile is $p^m=(\frac{1}{2}, \frac{1}{2})$. Example~1 is reminiscent of the introductory example in \cite{Tirole2016} in that there are no trading-up opportunities in the static monopoly outcome, such that no pricing dynamics emerge in the repeated game. 

Note that, for a durable variety, the supplier sets the price in the first period to $(1+\delta)/2$; for a rental variety, the supplier sets the price equal to $1/2$ in each period. That is, while the same net present value of the payment emerges, the consumer pays upfront for a durable variety and repeatedly for a rental variety. All consumer types buy in the first period, that is, $\F( V(\{o,p^1,a\})) =  \F( V(\{o,p^1,b\})) = 1/2$, and $\F( V(\{o,p^1,o\}))=0$.

For the second example, assume that consumer types are uniformly distributed on $V=\{v \in [0,1]^2 | v_a = v_b \}$, that is, $\F(E) = \int_E 1/\sqrt{2} d \mu(v)$, for $E \in \mathcal{B}(V)$, where $\mu$ is again the Lebesgue measure. Thus, $a$ and $b$ are perfect substitutes for all consumer types. We derive the following result in Appendix \ref{sec-online-appendix}, which is illustrated in the right panel of Figure~\ref{fig-example}.

\begin{example}[\textbf{Pricing dynamics}; $T=2$]\label{example-b}
 Let $x^0=o$ and assume that consumer types are uniformly distributed on $V=\{v \in [0,1]^2 | v_a = v_b \}$. Then, for two durables, two rentals, or mixed varieties, there exists a PBE where types $v\in V_1 = \{v|v_a=v_b \geq w \}$ are traded up in $t=1$, and types $v \in V_2 = \{v| w \geq v_a=v_b \geq p^2 \}$ are traded up in $t=2$, with cut-off value $w=(2+\delta)/(4+\delta)$ and $p^2=w/2$.
\end{example}

Regardless of the setting, the prices in period~2 are the same for first-time buyers/renters, and the same allocation emerges: The consumer buys the higher-valued variety within two periods or abstains from buying altogether. However, the prices in period~1 differ across settings: the price of a durable variety equals the present discounted value of the prices paid for a rental variety over two periods. Thus, the price of the durable variety includes future values; the price of the rental variety, by contrast, allows for a per-period treatment.

Formally, the price of the rental variety in period~1 is $p_i^1 =w(2-\delta) / 2 $, which is below the threshold type $w$: some consumer types strategically delay their consumption. Thus, there are trading-up opportunities for the seller and pricing dynamics emerge. 

Note that profit-maximizing prices of the static game are $p^m=(1/2, 1/2)$. Thus, half of the consumer types refrain from buying, resulting in a profit of $1/4$,  and leaving trading-up opportunities for types $v_i \leq 1/2$.

\section{Transitional games}\label{sec-ext-transitional}

In this section, we show how our analysis can be extended to settings in which one of the states is only indirectly accessible from the initial state via another state. We dub this class of settings ``transitional games.'' \autoref{fig-transitionalgame} provides an example of a transitional game with full support where absorbing variety $b$ is only indirectly accessible via rental variety $a$ from the initial state $x^0 = o$.
States that are only indirectly accessible from the initial state pose a challenge for our analysis, because the consumption options accessible in the static game do not correspond to those in the dynamic game, such that there might be no prices that exhaust all trading-up opportunities in the static game.

We now show that our approach of characterizing the dynamic equilibrium by analyzing the associated static game can nevertheless be applied to transitional games by introducing a suitably constructed ``extended static game.''

\begin{definition}[\textbf{Extended static game}]\label{def-extendedgame}
Consider a transitional game $(x^0=o, \Gamma, \F)$, where $b$ is only indirectly accessible via $a$, from the initial state $x^0 = o$. In the associated extended static game, let the probability measure of the consumer's valuation profile be the pushforward measure $\F^e(E) := \F(l^{-1}(E))$, for all $E \in \B(V^e)$, following from the linear transformation of the support $l: V \to V^e$, where
\begin{equation*}
l(v) = \begin{pmatrix}
1 & 0  \\
\frac{1}{\Delta} & \frac{\Delta-1}{\Delta}
\end{pmatrix} v, ~ v \in V\textbf{.}
\end{equation*}
Moreover, let the extended set of admissible transitions be $\Gamma^e$. 
\end{definition}

Thus, in the extended static game, the consumer chooses between the directly accessible states $\{x^0, a\}$ and a virtual state $\tilde{x}$ with value $\tilde{v} = v_{a} / \Delta +  v_{b} (\Delta-1)/\Delta$. The value of the virtual state corresponds to the average per period value if the consumer first chooses $a$ and $b$ afterwards, i.e., the shortest way to reach $b$.

Equivalent to our analysis above, let the static profit of the extended game be $\pi^e (p)= p_a \F^e(v \in V^e | v_a-p_a \geq \max\{v_b - p_b, 0 \}) + p_b \F^e(v \in V^e | v_b-p_b \geq \max\{v_a - p_a, 0 \})$ and denote the monopoly price by $p^{m,e} \in \argmax \pi^e(p)$. Moreover, let the set of price profiles that induce an allocation which leaves no trading-up opportunities for the seller in the extended static game be 
\begin{equation*}
  \Omega^e = \left\{p \in [\psi,1]^2 \Big| \F^e \left( \Big \{ v \in V^e (\{x^0, p^1, x^1 \}) ~\Big| x^{1} \notin \argmax_{x \in X(h^2)} v\cdot x \Big \} \right) =0 \right\}.
\end{equation*}

\begin{figure}[h]
\caption{A transitional game with associated extended static game}\label{fig-transitionalgame}
\begin{subfigure}{\textwidth}
	\vspace{-0.5em}
    	\begin{subfigure}[t]{0.29\textwidth}
    	\centering	
    	\caption{Transitional game}
    	\begin{tikzpicture}[thick, node distance = 2cm, shorten >= 3pt, shorten <= 3pt, ->]
    	\node[state, initial,inner sep=1pt,minimum size=25pt] (o) {$o$};
    	\node[state, left of=o, below of=o,inner sep=1pt,minimum size=25pt] (a) {$a$};
    	\node[state, right of=o, below of=o,inner sep=1pt,minimum size=25pt] (b) {$b$};
    	\draw (o) edge[bend left=15, right] node{$(o,a)$} (a);
    	\draw (a) edge[bend left=15, left] node{$(a,o)$} (o);
    	\draw (a) edge[below] node{$(a,b)$} (b);
    	\draw (a) edge[loop below  ] node{$(a,a)$} (a);
    	\draw (b) edge[loop below] node{$(b,b)$} (b);
    	\draw (o) edge[loop above] node{$(o,o)$} (o);
    	\end{tikzpicture}
    	\end{subfigure}
        \hspace{0.8cm}
     	\begin{subfigure}[t]{0.29\textwidth}
    	\centering	
    	\caption{Extended static game}
    	\begin{tikzpicture}[thick, node distance = 2cm, shorten >= 3pt, shorten <= 3pt, ->]
    	\node[state, initial,inner sep=1pt,minimum size=25pt] (o) {$o$};
    	\node[state, left of=o, below of=o,inner sep=1pt,minimum size=25pt] (a) {$a$};
    	\node[state, right of=o, below of=o,inner sep=1pt,minimum size=25pt, draw=blue] (b) {\textcolor{blue}{$\tilde{x}$}};
    	\draw (o) edge[bend left=15, right] node{$(o,a)$} (a);
    	\draw (a) edge[bend left=15, left] node{$(a,o)$} (o);
    	\draw (a) edge[below, blue] node{\textcolor{blue}{$(a,\tilde{x})$}} (b);
    	\draw (o) edge[right, blue] node{\textcolor{blue}{$(o,\tilde{x})$}} (b);
    	\draw (a) edge[loop below] node{$(a,a)$} (a);
    	\draw (b) edge[loop below, blue] node{\textcolor{blue}{$(\tilde{x},\tilde{x})$}} (b);
    	\draw (o) edge[loop above] node{$(o,o)$} (o);
    	\end{tikzpicture}
    	\end{subfigure}
        \hspace{0.6cm}
    	\begin{subfigure}[t]{0.25\textwidth}
    	\caption{Support}
    	\vspace{0.5cm}
        \centering
        \begin{tikzpicture} [scale=0.25]
        \fill[red,opacity=0.4] (-2.4,-2) -- (0,0) -- (10,0) -- (10,-2);
        \draw[line width=0.6mm,->] (-1,0) -- (11.75,0);
        \draw[line width=0.6mm,->] (0,-1) -- (0,11.75);
        \draw[line width=0.4mm] (10,0) -- (10,10) -- (0,10);
        \fill[blue,opacity=0.1] (0,0) -- (10,0) -- (10,10) -- (0,10);
        \node[text=blue] at (5,5) {$V$};
        \node at (-0.75,10) {$1$};
        \node at (10,-0.9) {$1$};
        \node at (11.25,-1.25) {$v_a$};
        \node at (-2.25,11.25) {$v_b, \textcolor{blue}{\tilde{v}}$};
        \draw[line width=0.4mm, blue, dashed] (0,6) -- (10,10);
        \draw[line width=0.4mm, blue, dashed] (0,0) -- (10,4);
        \draw[pen colour = {blue}, decorate, decoration = {calligraphic brace, mirror, raise = 2.5pt}, line width = 0.4mm] (10,4.2) --  (10,9.8);
        \node[text=blue] at (12.4,7) {$\frac{\Delta}{1-\Delta}$};
        \draw[pen colour = {blue}, decorate, decoration = {calligraphic brace, mirror, raise = 2.5pt}, line width = 0.4mm] (10,0.2) --  (10,3.8);
        \node[text=blue] at (11.7,2) {$\frac{1}{\Delta}$};
        \end{tikzpicture}
        \end{subfigure}
	\end{subfigure}
\end{figure}
The construction of the extended static game is illustrated in \autoref{fig-transitionalgame}. Panel~(a) shows the transitional game in its original form, where state $b$ is only indirectly accessible from the initial state $o$ via state $a$. Panel~(b) shows the extended static game where state $b$ is replaced by the virtual state $\tilde{x}$ (indicated in blue), which is accessible via the direct transition $(o,\tilde{x})$. Note that the value of the virtual option $\tilde{v}$ is different from $v_b$ for all off-diagonal value profiles in the support. The support of the value profiles $(v_a,\tilde{v}) \in V^e$ in the extended static game follows from the linear transformation $l$ of the original support $V$ and lies inside the dashed lines in panel (c).

We denote the price profile associated with the profit maximum that leaves no trading-up opportunities in the extended static game by $\bp^e$. Crucially, setting prices to zero in the extended static game leaves no trading-up opportunities, and hence $\bp^e$ always exists. We can then define a set of prices $\tilde{p} = (p^1_{a}, p^2_{a}, p^2_{b})$ that satisfy the following two restrictions
\al{
p_{a}^1 + (\Delta-1) p_{a}^2 &= \bp^e_{a} \Delta \text{,} \label{extented-game-price1}\\
p_{a}^1 + (\Delta-1) p_{b}^2 &= \bp^e_{\tilde{x}} \Delta  \label{extented-game-price2} \text{.}
}

The monopoly strategy for the seller, $\sigma^m$, then consists of playing prices $\tilde{p}$ in their respective states, i.e. $p^1_{a}$ in state $x^0$ and $p^2_{a}, p^2_{b}$ in states $a, b$ at any period and history, analogous to the seller playing the same static optimal prices repeatedly in the monopoly strategy defined in \autoref{sec-no-TUO}. We can then state the following result that shows how our approach of examining trading-up opportunities in static monopoly carries over to transitional games.

\begin{proposition}[\textbf{Transitional game}]\label{prop-4}
Consider a transitional game $(x^0=o, \Gamma, \F)$, and suppose there are no trading-up opportunities in the associated extended static monopoly outcome, that is, $p^{m,e}\in\Omega^e$. Then, in any PBE,
\begin{itemize}
    \item[(i)] the seller can do no better than follow the monopoly strategy along the equilibrium path.
    \item[(ii)] the seller's profit equals the (present discounted) sum of the repeated static monopoly profit of the extended static game, that is, $\Pi = \pi^e(p^{m,e}) \Delta$.
\end{itemize}
\end{proposition}

Proposition \ref{prop-4} implies that the key insights of our main analysis also apply to transitional games, including our approach of checking for trading-up opportunities in the associated (extended) static game in order to characterize the outcome of the repeated game. However, a subtle difference arises with transitional games: rather than excluding pricing dynamics, $\pi^e(p^{m,e}) = \pi^e(\bp^e)$ allows for a one-time price change in the dynamic game, which accounts for the change in available consumption options after the first period. Hence, the prices associated with the monopoly strategy of the seller, $\tilde{p}$, must differentiate between the first period and subsequent periods and allow for a one-time price change of consumption option $a$.

\autoref{fig-transitionalgame} illustrates a setting with a one-time price increase.
Just as in the setting with mixed varieties, the seller can set a price of zero for the virtual option~$\tilde{x}$ and a strictly positive price for the rental variety $a$, thereby ensuring that no trading-up opportunities are left while achieving a positive profit for the given support, $\pi^e(\bp^e) > 0$. But to implement this in the dynamic game, the seller must set a price of zero for variety $a$ in the first period, since playing a negative price for variety $b$ in future periods cannot constitute a PBE. In effect, the seller is accepting zero profit in the first period to move back from a transitional game to a setting where all consumer types can directly access both states.

\section{Conclusion}\label{sec-conclusion}

We have studied a unified analytical framework that captures a broad class of dynamic monopoly pricing problems, including multiple durable, multiple rental, or a mix of varieties. Our analysis demonstrates that the driving force behind pricing dynamics is the existence of trading-up opportunities. 

In particular, we show that the emergence of price dynamics hinges on whether the monopoly outcome in the corresponding static game leaves trading-up opportunities to the seller. If there are no trading-up opportunities, then the seller can implement the repeated static monopoly outcome regardless of commitment. Instead, with trading-up opportunities, the seller lowers prices to trade up positive masses of consumer types to higher-valued consumption options until all trading-up opportunities are exhausted or the game ends.  Our analysis shows that the essence of Coase's insight generalizes beyond the durable goods case: pricing dynamics emerge when the seller has an incentive to switch some consumer types to higher-valued consumption options.

Our findings imply that dynamic monopoly pricing problems can be analyzed by checking for trading-up opportunities in the static monopoly outcome. This approach also works for transitional games, where one of the varieties is only indirectly accessible from the initial state, provided that the associated extended static games are properly defined.

There is ample scope for future research. In particular, it would be interesting to study the seller's endogenous choice among alternative modes of trade.

\bibliographystyle{plainnat}
\bibliography{multi}

\newpage
\begin{appendix}
\section{Proofs}
\subsection{Proof of Lemma 1}
Since type $v$ obtains a higher total value along path $\x_k^t$ than along path $\x_l^t$ by assumption, we must have
\eqst{
  \nu(v,\x_k^t) - \rho(\p_k^t, \x_k^t)\geq \nu(v,\x_l^t) - \rho(\p_l^t, \x_l^t)\text{.}
}
Now, consider some type $\tilde{v} \neq v$. Then, we have
\alst{
 \nu(\tilde{v},\x_k^t) - \rho(\p_k^t, \x_k^t) &= \nu(v,\x_k^t) - \rho(\p_l^t, \x_l^t) + \nu(\tilde{v},\x_k^t)-\nu(v,\x_k^t)\\
&\geq  \nu(v,\x_l^t) - \rho(\p_l^t, \x_l^t) + \nu(\tilde{v},\x_k^t)-\nu(v,\x_k^t)\text{,}
}
since type $v$ obtains a higher total value along path $\x_k^t$ than along path $\x_l^t$ by assumption. For type $\tilde{v}$ to obtain a higher total value along path $\x_k^t$, it is thus sufficient to have
\eqst{
\nu(v,\x_l^t) - \rho(\p_l^t, \x_l^t) + \nu(\tilde{v},\x_k^t)-\nu(v,\x_k^t) \geq \nu(\tilde{v},\x_l^t) - \rho(\p_l^t, \x_l^t) \text{,}
}
which can be rearranged to yield the result in (\ref{equ-skim}).

\subsection{Proof of Proposition 1}

We prove the result by showing that (i) the maximum profit for the seller is the repeated static monopoly profit, and (ii) when facing prices $p^m$, buyers behave as if they were myopic, implying that the seller can do no better than following the monopoly strategy.

(i) Recall from (\ref{seller-profit}) that the seller's present discounted profit at history $h^t$ is 

\eq{\label{seller-profit-appendix}
 \Pi(h^t) =\sum_{\x^t\in \X(h^t)}\rho(\p^t, \x^t| h^t) \S \left( \x^t |h^t\right)\text{,}
}
where the strategy profile $\{\sigma,\hat{\sigma}\}$ is suppressed from the arguments for ease of notation. The following auxiliary result shows that $\Pi(h^t)$ can also be expressed in terms of the differences in the total values obtained by buyer types that are indifferent between different consumption paths. Letting
\eq{\label{equ-indifference}
 \Delta\nu(\x_k^t,\x_{k-1}^t | v_k)\equiv \nu(v_k,\x_k^t)-\nu(v_k,\x_{k-1}^t)=\rho(\p_k^t,\x_k^t)-\rho(\p_{k-1}^t,\x_{k-1}^t) \text{,}
}
denote the difference in the total values obtained by buyer types $v_k$ that are indifferent between consumption paths $\x_k^t$ with price path $\p^t_k$ and $\x_{k-1}^t$ with  $\p_{k-1}^t$, $k\geq 1$, if $v_k$ exists,\footnote{Note that at least one indifferent type $v_k$ exists for any consumption path with $\S(\x^t_k|h^t)>0$ if prices are profit maximizing.} we can restate the seller's present discounted profit at history $h^t$ as follows:

\begin{lemma}\label{lemma-rewrite}
The seller's present discounted profit at history $h^t$ with measure $\F(V(h^t))$ of active buyer types can be expressed as
\begin{equation*}
\Pi(h^t) = \rho(\p_0^t, \x_0^t |h^t) \F (V( h^t )) + \sum_{k\geq 1}^{} \Delta \nu (\x_k^t, \x_{k-1}^t| v_k) \sum_{j\geq k} \S(\x_j^t | h^t)\text{,}
\end{equation*}
where $\S(\x_j^t| h^t)$ is the measure of active buyer types that follow consumption path $\x_j$ after history $h^t$, and admissible consumption paths $\X(h^t)$ are ordered such that $(\X(h^t), \preceq)= \{\mathbf{x}_{0}^t,\mathbf{x}_{1}^t,... | h^t\}$.
\end{lemma}

\begin{proof}First, select any order $(\X(h^t), \preceq) = \{\mathbf{x}_{0}^t,\mathbf{x}_{1}^t,... | h^t\}$. Now, consider the first two paths $\x_{1}^t \neq \x_{0}^t$. Any indifferent type $v_1$ that obtains the same total payoff from both paths satisfies the indifference condition
\alst{
	\nu(v_1,\x_{1}^t)-&\rho(\p_{1}^t,\x_{1}^t) = \nu(v_1,\x_{0}^t)-\rho(\p_{0}^t,\x_{0}^t)
\intertext{or, equivalently,}
	&\rho(\p_{1}^t, \x_{1}^t) = \rho(\p_{0}^t,\x_{0}^t) + \Delta \nu (\x_1^t, \x_0^t| v_1) \text{.}
\intertext{Similarly, for paths $\x_2^t \neq \x_1^t$, we have}
    &\rho(\p_{2}^t,\x_{2}^t) = \rho(\p_{1}^t,\x_{1}^t) + \Delta \nu (\x_2^t, \x_1^t | v_2)
\intertext{and substituting from above yields}
    &\rho(\p_{2}^t, \x_{2}^t) = \rho(\p_{0}^t, \x_{0}^t) + \Delta \nu (\x_2^t, \x_1^t | v_2) + \Delta \nu (\x_1^t, \x_0^t | v_1) \text{.}
}
Iterating this procedure, for an arbitrary path $\x_k^t \in (\X(h^t), \preceq)$ we obtain
\eqst{
\rho(\p_{k}^t, \x_{k}^t) = \rho(\p_{0}^t, \x_{0}^t) + \sum_{1 \leq l \leq k}^{}\Delta \nu (\x_l^t, \x_{l-1}^t | v_l)\text{.}
}

Therefore, adding up the total payments made by active buyer types $\F(V(h^t)) = \sum_{k \geq 0} \S(\x_k^t| h^t)$ for order $(\X(h^t), \preceq)$, the present discounted profit at history $h^t$ satisfies
\eqst{
\Pi(h^t) = \rho(\p_0^t, \x_0^t |h^t) \F \left( V ( h^t ) \right) + \sum_{k\geq 1}^{} \Delta \nu (\x_k^t, \x_{k-1}^t | v_k) \sum_{j\geq k} \S(\x_j^t | h^t) \text{.}
}
\end{proof}

Now, suppose that both the seller and the buyer follow a monopoly strategy. Then, using Lemma \ref{lemma-rewrite}, the seller's profit from $t=1$ onward is given by
\al{\label{profit-simple}
\Pi(h^1) = \rho(\p^m, \x_a |h^1) + \Delta \nu (\x_b, \x_a | v) \S(\x_b| h^1) \text{,}
}
where $\p^m=\{p^m, p^m,...,p^m\}$, $\x_a=\{a,a,...,a\}$ and $\x_b=\{b,b,...,b\}$, respectively. Because there are no trading-up opportunities, we must have $\S(\x_k|h^1)=0$ for any path $\x_k\neq \x_i$ for which $\Delta\nu(\x_k,\x_i | v_k)>0$, $i=a,b$. Hence, the repeated static monopoly profit in (\ref{profit-simple}) is the maximum profit.

(ii)
First, observe that at any history $h^t$ all types choose their most-preferred variety when facing price profile $p^{\circ} = (-\Delta^{t},-\Delta^{t})$. To see this, note that types $v$ and $\tilde{v}$, $v \neq \tilde{v}$, can always mimic each other's behavior (i.e., make the same choices from $t$ onward), so that we have
\alst{
U(\tilde{v},x^t,h^t) - U(v,x^t,h^t) \leq \max_{i \in \{a,b\}} \{\tilde{v}_i - v_i\}\Delta^{t}, \ \ v \neq \tilde{v} \text{,}
}
where $U$ denotes the continuation valuation following choice $x^t$. Since the maximum value difference satisfies $\max_{i \in \{a,b\}} \{\tilde{v}_i - v_i\}=1$, all types purchase their most-preferred variety when facing prices $p^\circ$. In addition, it is straightforward that when facing $p^\circ$ in the static game, all types will equally accept and purchase their most-preferred variety so that $p^\circ \in \Omega$.

Now pick a price profile $\tilde{p}$ on the diagonal through the type space that satisfies
\alst{
\tilde{p} = (\min\{p^m_a, p^m_b\},\min\{p^m_a,p^m_b\}) - (\eta,\eta) \text{,}
}
for some $0\leq \eta \leq \Delta^{t}+\min\{p^m_a,p^m_b\}$. It is straightforward that all types $\max_{i \in \{a,b\}} \{v_i\} \geq \tilde{p}_i$ will accept and purchase their most-preferred variety when facing $\tilde{p}$ in the static game, implying that $\tilde{p} \in \Omega$ since $p^m \in \Omega$.

Denote by $x^{\circ}$ the choice that buyers make in the static game when facing prices $p^\circ$. We therefore have
\al{\label{eq-pcirc-stat}
&x^{\circ} \cdot (v-p^{\circ}) \geq x' \cdot (v-p^{\circ}), \ \ x^{\circ} \in \{a,b\}, x' \neq x^{\circ}, \ \ \forall \ v \in V \text{,}
\intertext{where we know that $x^\circ \in \{a,b\}$ as $p^\circ_a = p^\circ_b<0$. And, similarly, we also have that}
&x^{\circ} \cdot (v-p^{\circ}) + \delta U(v,x^{\circ},h^t) \nonumber \\
&\quad \quad \quad \quad \geq x' \cdot (v-p^{\circ}) + \delta U(v,x',h^t), \ \ x^{\circ} \in \{a,b\}, x' \neq x^{\circ}, \ \ \forall \ v \in V \text{.} \label{eq-pcirc}
}
By the definition of $p^{\circ}$ it then follows that
\al{\label{eq-pcirc-udiff-1}
\delta (U'-U^\circ) \leq (x^\circ-x') \cdot v\text{,} \ \ x^{\circ} \in \{a,b\}, x' \neq x^{\circ}, \ \ \forall \ v \in V \text{,}
}
where $U^{\circ}$ and $U'$ denote the continuation valuations associated with the choices $x^{\circ}$ and $x'$ respectively, given history $h^t$. This directly implies that
\alst{
x^{\circ} \cdot (v - &\tilde{p}) + \delta U(v,x^{\circ},h^t) \nonumber \\
&\geq x' \cdot (v-\tilde{p}) + \delta U(v,x',h^t), \ \ \ x^{\circ} \in \{a,b\}, x' \neq x^{\circ}, \ \ \forall \ v \in V \text{,}
}
that is, all types behave as if they were myopic when facing $\tilde{p}$ at any history $h^t$. This suffices to prove the statements if $p^m_a = p^m_b$.

To complete the proof for the case of $p^m_a \neq p^m_b$, consider a price profile
\al{\hat{p} = \left\{\begin{array}{ll} (\min\{p^m_a,p^m_b\},\min\{p^m_a,p^m_b\}) + (0,\varepsilon)  & \text{if } p^m_b>p^m_a \\
                                       (\min\{p^m_a,p^m_b\},\min\{p^m_a,p^m_b\}) + (\varepsilon,0) & \text{if } p^m_b<p^m_a
\end{array}
\right.
}
where $\varepsilon \in [0, \max\{p^m_a, p^m_b\} - \min\{p^m_a, p^m_b\} ]$.
Then by the same logic as above, we find that when facing $\hat{p}$ compared to $\tilde{p}$, the only types that now prefer the outside option to consumption also prefer the outside option at prices $p^m$ and the only types now preferring their non-most-preferred variety also do so at prices $p^m$. Thus, we find that all types follow the monopoly strategy at any history $h^t$ when facing prices $p^m$.

Then it follows that the seller can do no better than following the monopoly strategy along the equilibrium path, achieving the present discounted sum of the repeated static monopoly profit, as this is the highest possible profit as shown above.

\subsection{Proof of Lemma~\ref{cor1}}
Let $j$ be the variety with a strictly positive minimal value, i.e., $\underline{v}_j>0$. Suppose the seller sets prices $(p_a, p_b) = (\underline{v}_j, \underline{v}_j)$. Consider the static setup and note that with these prices all types prefer variety $j$ to the outside option $o$. Moreover, because prices are equal for both varieties, consumer types prefer variety $i$ to $j$ if and only if $v_i \geq v_j$. Therefore, consumer types buy their most-preferred variety if accessible and no trading-up opportunities remain, i.e., $(\underline{v}_j, \underline{v}_j) \in \Omega$.

Next, note that the seller obtains a profit of $\underline{v}_j$ in the static game with prices $(p_a, p_b) = (\underline{v}_j, \underline{v}_j)$, because every consumer type buys an accessible variety. Therefore, by definition $\pi(\bar{p}) \geq \underline{v}_j >0$.

\subsection{Proof of Proposition~\ref{prop-3}}

We prove the three statements in turn.

(i) Fix a PBE and consider a history $h^t$ with trading-up opportunities, such that $\F\left( \{ v \in V(h^t) ~| x^{t-1} \notin \argmax_{x \in X(h^t)} v\cdot x  \} \right) > 0$. Let $\bar{v}_i = \max_{v \in V(h^t)} v_i$ be the highest value for variety $i\in \{a,b\}$ of all types with this history, and let $\underline{v}_i = \min_{v \in V(h^t)} v_i$ be the lowest value. A type $v \in V(h^t)$ can be traded up at history $h^t$ if $v \cdot x^t > v \cdot x^{t-1}$ and $x^t \in X(h^t)$.  We denote the highest and lowest value of a type with history $h^t$ that can be traded up by $\bar{v}_i^{TU}$ and $\underline{v}_i^{TU}$, respectively.

Further denote the mass of types that are traded up at history $h^t$ by $\M^{TU}(h^t) = \F( \{ v\in V(h^t) ~| ~ v \cdot x^{t} > v \cdot x^{t-1}  \} ) $ and the remaining mass of types that are not traded up by $\M^{NTU}(h^t)$, such that $\F(V(h^t)) = \M^{TU}(h^t) + \M^{NTU}(h^t)$. We will show that, for any candidate equilibrium path  starting at history $h^t$ along which the seller does not trade up any types, trading up a positive measure of types instead is strictly profit-increasing. There are three cases to distinguish.

\textbf{Case 1}: $x^{t-1}=o$, and $\{a, b\} \cap X(h^t) \neq \emptyset$.\\
If $x^{t-1}$ is the outside option, the existence of trading-up opportunities implies that the seller can induce a positive measure of types $\M^{TU}(h^t)$ to buy variety $i \in \{a, b\}$ with $i \in X(h^t)$ at strictly positive prices, which is profit-increasing.   
%

\textbf{Case 2}: $x^{t-1} = j\in \{a,b\}$, $i\in X(h^t),i\neq j$, and $\bar{v}_i^{TU} > \bar{v}_j$.\\
For trading-up opportunities to exist, $x^{t-1}$ must be a non-absorbing variety $j$. Then the equilibrium profit from not trading up any types, $\hat{\Pi}(h^t)$, satisfies
\al{\label{prop2-profit-notradingup}
\hat{\Pi}(h^t) < \bar{v}_j \F(V(h^t))\Delta^t \text{,}
}
as the seller cannot extract the full surplus from all types $v$ at history $h^t$ with a linear price. If the seller trades up the mass of types $\M^{TU}(h^t)$, then the equilibrium profit obtained from types traded up, $\Pi^\ast (h^t)$, satisfies
\al{\label{prop2-profit-tradingup}
\Pi^\ast (h^t) \geq v_i^\ast \M^{TU}(h^t)\Delta^t \text{,}
}
where $v_i^\ast$ denotes the lowest value $v_i$ of the cutoff types who are indifferent to trading up to $i$, as the seller can always obtain at least the value of the lowest type in the set. The equilibrium profit obtained from types not traded up, $\Pi^\circ (h^t)$, satisfies
\al{\label{prop2-profit-remaining}
\Pi^\circ (h^t) <\bar{v}_j \M^{NTU}(h^t) \Delta^t \text{,}
}
because as before the seller cannot extract the full surplus using a linear price. As $\bar{v}_i^{TU} > \bar{v}_j$ by assumption, there exists a $v^\ast_i$ that satisfies $\bar{v}_i^{TU} > v_i^\ast > \bar{v}_j$. Therefore, (\ref{prop2-profit-notradingup}), (\ref{prop2-profit-tradingup}), (\ref{prop2-profit-remaining}), and $\F(V(h^t)) = \M^{TU}(h^t) + \M^{NTU}(h^t)$ together imply that
\alst{
\Pi^{*}(h^t) + \Pi^{\circ}(h^t) > \hat{\Pi}(h^t)\text{.}
}

\textbf{Case 3}: $x^{t-1} = j\in \{a,b\}$, $i\in X(h^t),i\neq j$, and $\bar{v}_i^{TU} < \bar{v}_j$.\\
Starting at $x^{t-1}=j$, there are three paths of play without trading up any types to $i$. We consider these in turn and show that in each case, there are deviation incentives for the seller, that is, trading up (some) types to $i$ is strictly profit increasing.

(a) Suppose the buyer always plays $j$ along the equilibrium path. As all types only ever accept at a price at which they obtain a (weakly) positive utility along the path of play, we must have that the equilibrium profit obtained satisfies
\al{
\Pi(h^t) \leq \underline{v}_j \F(V(h^t)) \Delta^t \text{.}
}
As $\bar{v}_i^{TU} > \underline{v}_j$ by the assumption that trading-up opportunities exist, there exists a $v_i^\ast$ that satisfies $v_i^\ast > \underline{v}_j$, which by (\ref{prop2-profit-tradingup}) implies that trading up (some) types to $i$ is strictly profit increasing.

(b) Suppose the seller induces some types to play $x^t = o$ and the outside option is non-absorbing. Consider the types $v$ for which $v_i > v_j$. Either, (some of) these types play $x^{t+1}=j$, such that case (a) applies at the continuation history for these types at time $t+1$, or they play $x^{t+1} = o$, such that Case 1 applies at $t+1$.
Trading up is strictly profit-increasing in either of these cases.

(c) Suppose the seller induces some types to play $x^t = o$ and the outside option is absorbing. As before, if there exist $v$ that satisfy $v_i > v_j$ that play $x^{t+1} = j$, then case (b) applies. If not however, all types $v$ for which $v_i > v_j$ must have played $x^t=o$ and can no longer be traded up, as the outside option is absorbing. Consider that types with history $h^t$ that satisfy $v_i > v_j$, must have faced some $p_j^{t-1} \leq \underline{v}_j^{TU}$ at time $t-1$, because in equilibrium, no type accepts unless they obtain a weakly positive present-discounted utility. But if at time $t$ the profit-maximizing choice for the seller is to play some $p_j^\ast \geq \bar{v}_j^{TU}$ at history $h^t$, then at time $t-1$ playing $p_j^{t-1}\leq \underline{v}_j^{TU}$ cannot be profit-maximizing for types with history $h^t$. Suppose the seller had instead played $p^\ast$ at time $t-1$. If this is the profit-maximizing price at $h^t$, then profit from the consumer types with history $h^t$ at time $t-1$ must be increased and the present-discounted total profit at time $t-1$ must also be increased, since all types with history $h^t$ that do not play $x^{t-1}=j$ at this higher price either play $x^{t-1} = o$, leaving overall profit unaffected, or play $x^{t-1}=i$ yielding a strictly positive profit. 

Then in conjunction statement (i) follows.

(ii) The proof follows the same lines as part (ii) of the proof of Proposition 1. We apply this proof structure to the set of all price profiles that leave no trading up opportunities in the static game, $\Omega$. Specifically, denote by $\Lambda$ the set of price profiles $p$ that leave no trading-up opportunities for any history $h^t$ in the dynamic game. We will now show that $\Omega \setminus \Lambda = \emptyset$ and $\bp \in \Lambda$. 

First, note that because $\bp \in \Omega$ by assumption, for a price profile $\tilde{p}$ on the diagonal through the type space, with $\eta \geq 0$, we have
\al{
\tilde{p} = (\min\{\bp_a,\bp_b\},\min\{\bp_a,\bp_b\}) - (\eta,\eta) \implies \tilde{p} \in \Omega \text{,} \label{eq-tildep-omega1}
\intertext{as all types willing to purchase at prices $\tilde{p}$ choose their most-preferred variety, and all types choosing the outside option will also do so at prices $\bp$. Similarly, for a price profile $\tilde{\tilde{p}}$ on the (vertical or horizontal) line between $\bar{p}$ and the diagonal, with $\eta \in [0,\max\{\bp_a,\bp_b\}-\min\{\bp_a,\bp_b\}]$, we have}
\tilde{\tilde{p}} = \left\{\begin{array}{ll}
     (\bp_a,\bp_b)-(0,\eta), &  \text{if } \bp_b>\bp_a\\
     (\bp_a,\bp_b)-(\eta,0), &  \text{if } \bp_b<\bp_a
\end{array}\right.
\implies \tilde{\tilde{p}} \in \Omega \text{,} \label{eq-tildep-omega2}
}
as all types purchasing a different variety at prices $\tilde{\tilde{p}}$ than at prices $\bp$ must now choose their most-preferred variety, and all types switching from the outside option to consumption must now choose their most-preferred variety. 

Second, recall from part (ii) of the proof of Proposition 1 that the price profile $p^{\circ} = (-\Delta^{t},-\Delta^{t})$ satisfies $p^{\circ} \in \Lambda$ as all types can always mimic each other's behavior. In addition, by (\ref{eq-tildep-omega1}) we also have that $p^\circ \in \Omega$.

Now pick a price profile $\hat{p}$ that satisfies $\hat{p} = p^{\circ} + (\varepsilon,\varepsilon)$ for some $\varepsilon \in [0 , \Delta^{t}+\min\{\bp_a,\bp_b\}]$. By (\ref{eq-tildep-omega1}) we know $\hat{p} \in \Omega$. Denote by $x^{\circ}$ the choice that buyers make in the static game when facing prices $p^\circ$. By (\ref{eq-tildep-omega1}) we then have
\al{\label{eq-pcirc-stat}
&x^{\circ} \cdot (v-p^{\circ}) \geq x' \cdot (v-p^{\circ}), \ \ x^{\circ} \in \{a,b\}, x' \neq x^{\circ}, \ \ \forall \ v \in V \text{,}
\intertext{where we know that $x^\circ \in \{a,b\}$ as $p^\circ_a = p^\circ_b<0$. Since $p^{\circ} \in \Lambda$, we also have that}
&x^{\circ} \cdot (v-p^{\circ}) + \delta U(v,x^{\circ},h^t) \nonumber \\
&\quad \quad \quad \quad \geq x' \cdot (v-p^{\circ}) + \delta U(v,x',h^t), \ \ x^{\circ} \in \{a,b\}, x' \neq x^{\circ}, \ \ \forall \ v \in V \text{,} \label{eq-pcirc}
}
By the definition of $p^{\circ}$ and (\ref{eq-pcirc}) it then follows that
\al{\label{eq-pcirc-udiff-1}
\delta (U'-U^\circ) \leq (x^\circ-x') \cdot v\text{,} \ \ x^{\circ} \in \{a,b\}, x' \neq x^{\circ}, \ \ \forall \ v \in V \text{,}
}
where $U^{\circ}$ and $U'$ denote the continuation valuations associated with the choices $x^{\circ}$ and $x'$ respectively, given history $h^t$. This also implies that
\al{
x^{\circ} \cdot (&v - p^{\circ}-\varepsilon) + \delta U(v,x^{\circ},h^t) \nonumber \\
&\geq x' \cdot (v-p^{\circ}-\varepsilon) + \delta U(v,x',h^t), \ \ \ \ x^{\circ} \in \{a,b\}, x' \neq x^{\circ}, \ \ \forall \ v \in V \text{.} \label{eq-pcirc-epsilon1}
}
Thus, for any $\varepsilon \in [0, \Delta^{t}+\min\{\bp_a,\bp_b\}]$, only types $v< \min\{\bp_a,\bp_b\}$ may choose $o$ over $x^\circ$ at prices $\hat{p}$, which continues to leave no trading-up opportunities since $\bp \in \Omega$ by assumption, and therefore $\hat{p} \in \Lambda$. Hence, for any $\tilde{p}$ that satisfies (\ref{eq-tildep-omega1}) we have $\tilde{p} \in \Lambda$.

Now fix the price profile $\hat{p} = (\min\{\bp_a,\bp_b\},\min\{\bp_a,\bp_b\})$. By (\ref{eq-tildep-omega2}) we have $\hat{p} \in \Omega$, and as shown above we also have $\hat{p} \in \Lambda$. Consider a price profile $p' = \hat{p} + (0,\varepsilon)$ if $\bp_b>\bp_a$ and $p' = \hat{p} + (\varepsilon,0)$ if $\bp_b<\bp_a$ where $\varepsilon \in [0,\max\{\bp_a,\bp_b\}-\min\{\bp_a,\bp_b\}]$. Then by the same logic as above, for any $\varepsilon \in [0, \max\{\bp_a,\bp_b\}-\min\{\bp_a,\bp_b\}]$, we find that the only types that may choose the outside option over consumption also choose the outside option at prices $\bp$ and the only types who may choose the other variety also do so at prices $\bp$. Thus, we find $p' \in \Lambda$ or equally that any $\tilde{p}$ that satisfies (\ref{eq-tildep-omega2}) satisfies $\tilde{p} \in \Lambda$ and therefore $\bp \in \Lambda$.

Finally, note that we can construct (\ref{eq-tildep-omega1}) and (\ref{eq-tildep-omega2}) for any price profile $p \in \Omega$ and thus we find that $\Omega \setminus \Lambda = \emptyset$. Then it follows from the definition of $\bp$ that the seller will never play prices below $\bp$, implying that the present discounted stream of profits $\pi(\bp)\Delta$ is a lower bound on the sellers' profit.


(iii)  Fix a PBE and consider a history $h^t$ with trading-up opportunities, that is, $\F\left( \{ v \in V(h^t) ~| x^{t-1} \notin \argmax_{x \in X(h^t)} v\cdot x  \} \right) > 0$. Let $\bar{v}_i(h^t) = \max_{v \in V(h^t)} v_i$ be the highest value for variety $i\in \{a,b\}$ of all types with this history $h^t$ and $\underline{v}_i(h^t) = \min_{v \in V(h^t)} v_i$ be the lowest value. A type $v \in V(h^t)$ can be traded up at history $h^t$ if $v \cdot x^t > v \cdot x^{t-1}$ and $x^t \in X(h^t)$.  We denote the highest and lowest value of a type with history $h^t$ that can be traded up by $\bar{v}_i^{TU}(h^t)$ and $\underline{v}_i^{TU}(h^t)$, respectively. We define analogously $\bar{v}_j^{TU}(h^t), \underline{v}_j^{TU}(h^t)$ for $j \in \{a,b\}, j \neq i$, if trading-up opportunities exist for $j$ as well. Further denote the mass of types that can be traded up by $\M^{TU}(h^t) = \F( \{ v\in V(h^t) ~| x^{t-1} \notin \argmax_{x \in X(h^t)} v\cdot x \} )$.

Let $\bar{v}^{TU}(h^t) = \max\{\bar{v}_i^{TU}(h^t), \bar{v}_j^{TU}(h^t)\}$ and $\underline{v}^{TU}(h^t) = \min \{\underline{v}_i^{TU}(h^t), \underline{v}_j^{TU}(h^t) \}$ denote the highest and lowest value, respectively, of the varieties that consumers at history $h^t$ can be traded up to. Assume without loss of generality that $\underline{v}_i^{TU}(h^t) \leq \underline{v}_j^{TU}(h^t)$ and consider some $\varepsilon(h^t)$ that satisfies
\alst{
\varepsilon(h^t) \geq \bar{v}^{TU}(h^t) - \underline{v}_i^{TU}(h^t) \text{.}
}
As the seller trades up a positive measure of consumers at any history $h^t$ with trading-up opportunities (see part (i)), by definition of $\bar{v}^{TU}(h^t)$ we have that $\bar{v}^{TU}(h^t) - \underline{v}_i^{TU}(h^t)$ must decrease with the length of a history by Lemma \ref{lemma-skimming}, such that a smaller $\varepsilon(h^t)$ will satisfy the above condition.\footnote{Lemma \ref{lemma-skimming} implies $\bar{v}^{TU}(h^t) > \bar{v}^{TU}(h^{t+1})$ because types with a high valuation are traded up earlier than types with a low valuation. Thus, with $T$ sufficiently large, $\varepsilon$ becomes sufficiently small over time.} We now show that for $\varepsilon(h^t)$ small enough, the seller strictly prefers to trade up all types at once if the minimal value of at least one variety and $\pi(\bp)$ are strictly positive. To ease notation, we henceforth suppress the conditioning of $\M^{TU}$, $\varepsilon$, $\bar{v}^{TU}$ and $\underline{v}_i^{TU}$ on history $h^t$ whenever possible.

As trading up will occur along the equilibrium path for any history with trading-up opportunities (see part (i)), consider $t$ to be the period at which trading-up is profit increasing for the seller for the given history. Let $\Pi^\ast(h^t)$ denote the equilibrium profit for the seller obtained from trading up only part of the mass $\M^{TU}$. As the seller cannot extract the full surplus with a linear price or trade up the remaining types before time $t+1$, there exists a $\lambda \in (0,1)$ such that
\alst{
\Pi^*(h^t) < \lambda \M^{TU} \bar{v}^{TU} \Delta^t + \delta (1-\lambda) \M^{TU}\bar{v}^{TU} \Delta^{t+1} \text{.}
}
In addition, let $\bar{\Pi}(h^t)$ denote the seller's equilibrium profit obtained from trading up all types. As the seller can always obtain at least the minimal value of a variety in each period, we have that
\al{
\bar{\Pi}(h^t) \geq  (1-\varphi) \M^{TU} \underline{v}_i^{TU} \Delta^t + \varphi \M^{TU} \underline{v}_j^{TU} \Delta^t \text{,}
\label{equation23}
}
with $\varphi \in [0,1]$. 

Using these profits and noting that $\delta \Delta^{t+1}=\Delta^t-1$, we can write
\alst{
\Pi^*(h^t) - \bar{\Pi}(h^t) &<  \lambda \M^{TU} \bar{v}^{TU} \Delta^t + \delta (1-\lambda) \M^{TU} \bar{v}^{TU} \Delta^{t+1} - \\ & \hspace{1.5 em} (1-\varphi) \M^{TU} \underline{v}_i^{TU} \Delta^t - \varphi \M^{TU} \underline{v}_j^{TU} \Delta^t \\
&=\left[(\Delta^t - 1 + \lambda) \bar{v}^{TU} - (1-\varphi) \underline{v}_i^{TU} \Delta^t  - \varphi \underline{v}_j^{TU} \Delta^t \right] \M^{TU}\\
&\leq \left[(\Delta^t - 1 + \lambda) (\varepsilon + \underline{v}_i^{TU}) - (1-\varphi) \underline{v}_i^{TU} \Delta^t  - \varphi \underline{v}_j^{TU} \Delta^t \right] \M^{TU} \\
&= \left[(\Delta^t - 1 + \lambda)\varepsilon - (1-\lambda) \underline{v}_i^{TU} + \varphi \Delta^t (\underline{v}_i^{TU}-\underline{v}_j^{TU})\right]\M^{TU} \text{.}
}
Therefore, $\bar{\Pi}(h^t) > \Pi^*(h^t)$ whenever
$$
\varepsilon (h^t) \leq \frac{ (1-\lambda)\underline{v}_i^{TU}(h^t) + \varphi \Delta^t (\underline{v}_j^{TU}(h^t)-\underline{v}_i^{TU}(h^t))}{\Delta^t-1+\lambda} \text{.}
$$
Recall that $\underline{v}_i^{TU}(h^t) \leq \underline{v}_j^{TU}(h^t)$ and note that the right-hand side is strictly positive if $\underline{v}_i^{TU}(h^t)>0$.

It remains to be shown that the right-hand side is also strictly positive if $\underline{v}_i^{TU}(h^t)=0$. To see this, note that $\underline{v}_j^{TU}(h^t)>0$ because the minimal value of at least one variety is strictly positive by assumption. Hence, it suffices to show that $\varphi>0$. Suppose instead that $\varphi=0$. From equation (\ref{equation23}) we get that $\bar{\Pi}(h^t)=0$, which is a contradiction: by Lemma \ref{cor1}, we know that $\underline{v}_j(h^t)>0$ implies $\pi(\bar{p})>0$, from which $\bar{\Pi}(h^t)>0$ follows by (ii).

\subsection{Proof of Proposition~\ref{prop-4}}

We first show that, for any strategy profile $\{\sigma, \hs\}$ of the dynamic game, we can define an associated static game that delivers the same payoffs to all players when multiplied with the (present discounted) number of periods $\Delta$, with the extended static game of Definition~\ref{def-extendedgame} as a special case. Recall from (\ref{seller-profit}) that the seller's profit at history $h^1$ is
\alst{
 \Pi(\sigma,\hs | h^1) =\sum_{\x^1\in \X(h^1)}\rho(\p^1, \x^1|\sigma,\hs, h^1) \S \left( \x^1 | \sigma,\hs,h^1\right) \text{,}
}
where $\rho(\p^1, \x^1|\sigma,\hs, h^1)$ is the strategy-contingent total payment along path $\x^1 \in \X^1(h^1)$ and $\S \left( \x^1 | \sigma,\hs,h^1\right)$ denotes the associated measure of types.
Now, let $\bar{\rho}(\x^1)$ denote the per-period payment that, if it were received in every period, would give the seller the same (present-discounted) total profit as consumption path $\x^1$,
\alst{
\bar{\rho}(\x^1)  = \frac{\rho(\p^1, \x^1|\sigma,\hs, h^1)}{\Delta}.
}
Similarly, let $\bnu(v,\x^1)$ be the per-period consumption value that, if it were received in every period, would give a consumer of type $v$ the same (present-discounted) total value as that obtained along path $\x^1$,
\alst{
\bnu(v,\x^1)  = \frac{\nu(v,\x^1)}{\Delta}.
}
Then, the seller's profit can be written as
\alst{
\Pi(\sigma,\hs | h^1) &=  \Delta \sum_{\x^1\in \X^1(h^1)} \bar{\rho}(\x^1) \S \left( \x^1 | \sigma,\hs,h^1\right) \text{,}
}
whereas the net value obtained by a type-$v$ consumer along path $\x^1$ becomes $\Delta (\bar{\nu}(v, \x^1) - \bar{\rho}(\x^1))$. That is, in the dynamic game associated with strategy profile $\{\sigma, \hs\}$, the seller obtains a fixed per-period payment along all admissible paths $\x^1$, multiplied by the (present discounted) number of periods $\Delta$. Each path $\x^1\notin \{\x_o, \x_a, \x_b\}$ represents a mixed consumption option.

Now consider the transitional game. Denote the monopoly prices in the associated extended static game by $p^{m,e}=(p_{a}^{m,e}, p_{\tilde{x}}^m)$. By Proposition \ref{prop-1} (i), we obtain that this is the maximum profit if prices $p^{m,e}$ leave no trading-up opportunities, or $\pi^e(p^{m,e})\Delta= \pi^e(\bp^e)\Delta$. 

Assume that this is the case, that is, $p^{m,e} \in \Omega^e$ and consider the monopoly strategy for the seller $\sigma^m$ in the transitional game that consists of playing the tuple of prices $\tilde{p}=(p_{a}^1,p_{a}^2, p_{b}^2)$ repeatedly in the associated states. By (\ref{extented-game-price1}) and (\ref{extented-game-price2}), this strategy yields the maximum profit for the seller if the buyer plays the monopoly strategy. Now observe that (\ref{extented-game-price1}) and (\ref{extented-game-price2}) also imply that
\al{\label{indifference-equivalence}
v_{a} - p^2_{a} \geq v_{b} - p^2_{b}  \implies  v_{a} - \bp_{a}^e \geq \tilde{v} - \bp_{\tilde{x}}^e
\text{,}
}
which can be seen by substituting $p^2_{a}, p^2_{b}$ from the definitions of $\bp^e_{a}$ and $\bp^e_{\tilde{x}}$ and $\tilde{v}$ into the indifference condition. That is, prices $p^2_{a},p^2_{b}$ implement the same indifference condition in the one-shot game in any period $t>1$ in the transitional game at states $a, b$ as prices $\bp^e_{a}, \bp^e_{\tilde{x}}$ do in the extended static game. 

As the outside option is non-absorbing, it must be that no positive measure of types is allocated to the outside option when facing prices $\bp^e$ in the extended static game. Similarly $a$ is non-absorbing by construction and thus no positive measure of types that prefer $b$ must be allocated to $a$ at any time $t>1$ and history $h^t$. By (\ref{indifference-equivalence}) therefore, if no positive measure of types is allocated to the outside option in the transitional game at time $t=1$, then at time $t=2$ and history $h^{2}$ strategy $\sigma^m$ implements the same allocation in the one-shot game of the transitional game as prices $\bp^e$ do in the extended static game and thus leaves no trading-up opportunities. We can therefore directly apply part (ii) of the proof of Proposition \ref{prop-1}  to prove that at any history of the transitional game at any time $t>1$ and states $a, b$, the resulting allocation when playing prices $(p^2_{a}, p^2_{b})$ will (i) leave no trading-up opportunities and (ii) achieve the maximum profit if no positive measure was allocated to the outside option at time $t=1$.

Thus it only remains to check that playing $\sigma^m$ also ensures that no positive measure of types is allocated to the outside option at $t=1$, or
\al{\label{indifference-transitional-period1}
(v - p^1_{a}) \cdot a + \delta U(v,a, h^1) \geq (v - 0) \cdot o + \delta U(v,o,h^1) \ \forall v \in V \text{,}
}
where $U$ denotes the continuation utility following $a$ and $o$ respectively. Given strategy $\tilde{\sigma}$, it is straightforward that $U(v,a,h^1) \geq 0 \ \forall v \in V$ by construction. We therefore find that playing $p_{a}^1 \leq \underline{v}_{a}$ ensures that (\ref{indifference-transitional-period1}) is satisfied, where $\underline{v}_{a} = \min_{v \in V} v_a$. Thus, the seller can play $\tilde{\sigma}$ and all types will allocate themselves as they do in $\pi^e(\bp^e)$, while prices satisfy the necessary restrictions to ensure the sellers profit is $\Delta \pi^e(\bp^e)$.

\newpage
\section{Examples}\label{sec-online-appendix}


\subsection{Example \ref{example-a}: Repeated static monopoly }

Let $V=\{v \in [0,1]^2 | v_b = 1- v_a \}$, as illustrated in Figure \ref{fig-example} and $\F(E) = \int_E 1/\sqrt{2} d \mu(v)$, for $E \in \mathcal{B}(V)$, where $\mu$ is the Lebesgue measure.

Let us start with the static monopoly prices. Note that any price profile with $p_a+p_b<1$ yields a lower profit than $p_a+p_b=1$. For $p_a+p_b \geq 1$ the demand function for variety $i \in \{a,b\}$ is $d_i(p)=1-p_i$; thus, we obtain $\pi(p)=\sum_i (1-p_i)p_i$ and the resulting static monopoly price profile is $p^m = (1/2,1/2)$. 

Next, consider the set of price profiles that leave no trading-up opportunities in the static game. (i) for two durable varieties, $\Omega = \{ p \in [\psi,1]^2 | p_a + p_b  \leq 1 \}$; (ii) for two rental varieties, $\Omega = \{ p \in [\psi,1]^2 | p_a = p_b \leq 1/2 \}$; (iii) and for mixed varieties, where $a$ is the rental and $b$ is the durable, $\Omega = \{ p \in [\psi,1]^2 | p_b \leq \min\{1- p_a, p_a \} \}$. We obtain in all three cases $p^m \in \Omega$, thus, no trading-up opportunities exist at the static monopoly price profile. 

Next, we derive a PBE for $T=2$. Let us start with (i) two durables. Suppose only consumer types $v_i \leq w_i, i \in \{a,b\}$, remain in the market in period~$2$.\footnote{Consumer type $v=(v_a,v_b)$ buys variety $i$ in period~1 iff $v_i + \delta v_i - p_i^1 \geq \max \{ \delta(v_i - p_i^2), v_j + \delta v_j - p_j^1 ,\delta(v_j - p_j^2), 0) \} $. In the symmetric equilibrium, which we are constructing, this condition simplifies to $v_i \geq w_i$.} Any price profile $p_a^2 + p_b^2 <1$ is not profit maximizing.\footnote{The supplier can increase prices to $p+(\varepsilon, \varepsilon)$, for $\varepsilon \in (0, (1-p_a-p_b)/2]$, without affecting the demand resulting in a higher profit.} The demand function in period~$2$ is, therefore, $d_i(p_i^2) = (w_i - p_i^2)$. Maximizing the profit in the second period subject to $p_a^2 + p_b^2 \geq 1$ yields $p_i^2 = 1/2 + (w_i - w_j) / 4$, with $i \neq j \in \{a,b\}$. Next, let $w_i = p_i^1 - \delta p_i^2$ be the cut-off type who strategically delays his purchase.\footnote{In a symmetric equilibrium, consumer types purchase their higher-valued variety or none. Therefore, the cut-off type is defined as being indifferent between buying the same variety today or tomorrow.} The discounted sum of profits is $\sum_{i \in \{a,b\} } (1-w_i)p_i^1 + \delta \sum_{i \in \{a,b\} } (w_i-p_i^2)p_i^2$. Note that the cut-off types determine all prices, thus, we can rewrite the supplier's profit function in terms of the cut-off types $(w_a,w_b)$, which determine when to trade-up which types:
$$
\sum_{i \in \{a,b\} } (1-w_i) \left( w_i + \delta \left ( \frac{1}{2} + \frac{w_i-w_j}{4} \right) \right) + \delta \sum_{i \in \{a,b\} } \frac{3w_i +w_j -2}{4} \left ( \frac{1}{2} + \frac{w_i-w_j}{4} \right).
$$
The resulting profit maximizing cut-off types are $w_i= 1/2$, implying $p_i^2=1/2$ and $p_i^1 = (1+\delta)/2$. Thus, all consumer types buy in $t=1$ their higher-valued variety. 

Next, consider (ii) two rental varieties. The above analysis implies that the price for first-time renters in period~$2$ is $p_i^2(o) =  1/2 + (w_i - w_j) / 4$. The profit that the supplier obtains from these types is thus equal to the profit that the supplier in (i) obtains from buyer types in the second period. The cut-off type is $w_i = p_i^1 + \delta p_i^2(i) - \delta p_i^2(o)$,\footnote{Consumer type $v=(v_a,v_b)$ buys in the first period iff $v_i + \delta v_i - p_i^1 - \delta p_i^2(i) \geq \max \{ \delta(v_i - p_i^2(o)), v_j + \delta v_j - p_j^1 - \delta p_j^2(j),\delta(v_j - p_j^2(o)), 0, v_j - p_j^1 + \delta v_i - \delta p_i^2(j), v_i - p_i^1 + \delta v_j - \delta p_j^2(i) \}. $} and the net present value of the price for repeat renters is $p_i^1 + \delta p_i^2(i)$. The supplier obtains the profit $\sum_{i \in \{a,b\} } (1-w_i)(p_i^1 + \delta p_i^2(i)) + \delta \sum_{i \in \{a,b\} } (w_i-p_i^2)p_i^2$, which is equivalent to the profit in (i) after plugging in for $w_i$. Therefore, $w_i=1/2$, implying $p_i^2(o)=1/2$ and $p_i^1 + \delta p_i^2(i) = (1+\delta)/2$. By positive selection, we get $p_i^2(i)=w_i =1/2$ and hence $p_i^1=1/2$.\footnote{The price for switchers or non-switchers $p_i(j)$ is set such that the cut-off type is indeed $w_i$.} Prices are constant and all types buy their higher-valued variety in $t=1$. 

Finally, consider (iii) mixed varieties. Note that in period~$2$, there is no difference between the varieties: every decision is final. The supplier's profit for product $a$ takes the same form as in (ii); the supplier's profit for product $b$ takes the same form as in (i). Thus, we obtain $w_i=1/2$ as the profit maximizing cut-off values. Prices for variety $a$ are given by (ii) and prices for variety $b$ are given by (i). All consumer types buy their higher-valued variety in $t=1$. 

\subsection{Example \ref{example-b}: Pricing dynamics}

Let $V=\{v \in [0,1]^2 | v_b = v_a \}$, as illustrated in Figure \ref{fig-example} and $\F(E) = \int_E 1/\sqrt{2} d \mu(v)$, for $E \in \mathcal{B}(V)$, where $\mu$ is the Lebesgue measure.

Let us start with the static monopoly prices. Since $a$ and $b$ are perfect substitutes, only the price $p_l = \min \{p_a, p_b\}$ matters, and the demand function $d(p)=1-p_l$ results. Maximizing the profit $\pi(p)=(1-p_l)p_l$ yields the the monopoly price $p_l^m =1/2$. Therefore, $p^m \in \{p \in [\psi,1]^2 | \min\{p_a,p_b\} = 1/2 \}$ is the set of (outcome-equivalent) price profiles that maximize static profit. 

Next, consider the set of price profiles that leave no trading-up opportunities in the static game. (i) for two durable varieties, $\Omega = \{ p \in [\psi,1]^2 | \min\{p_a , p_b\}  \leq 0 \}$; (ii) for two rental varieties, $\Omega = \{ p \in [\psi,1]^2 | p_a = p_b \leq 0 \}$; (iii) and for mixed varieties, where $a$ is the rental and $b$ is the durable, $\Omega = \{ p \in [\psi,1]^2 | p_b \leq \min\{0, p_a \} \}$. We obtain in all three cases $p^m \notin \Omega$. Therefore, there exist trading-up opportunities at any static monopoly price profile.

Next, we derive a PBE for $T=2$. Let us start with (i) two durables. Suppose only consumer types $v_i \leq w, i \in \{a,b\}$ remain in the market in period~$2$.\footnote{Consumer type $v=(v_a,v_b)$ buys variety $i$ in period~1 iff $v_i + \delta v_i - p_i^1 \geq \max \{ \delta(v_i - p_i^2), v_j + \delta v_j - p_j^1 ,\delta(v_i - p_i^2), 0) \} $. In the equilibrium, which we are constructing, this condition simplifies to $v_i \geq w$.} The demand function in period~$2$ is $(w - p_l^2)$, where $p_l^2= \min\{p_a^2,p_b^2\}$. Maximizing the profit in the second period results in $p_l^2 = w/2$. Next, let $w = p_l^1 - \delta p_l^2$ be the cut-off type who strategically delays his purchase.\footnote{Because $a$ and $b$ are perfect substitutes, the cut-off type is defined as being indifferent between buying a variety today or tomorrow.} The discounted sum of profits is $(1-w)p_l^1 + \delta (w-p_l^2)p_l^2$. Note that the cut-off types determine all prices, thus, we can rewrite the supplier's function in terms of the cut-off value $w$, which determines when to trade-up which types:
$$
(1-w)\left( 1+ \frac{\delta}{2} \right) w + \delta \left(w- \frac{w}{2} \right) \frac{w}{2} .
$$
The resulting profit maximizing cut-off type is $w= (2+\delta)/(4+\delta)$, implying $p_i^2=(2+\delta)/(8+2\delta)$ and $p_i^1 = (2+\delta)^2/(8+2\delta)$. Thus, pricing dynamics emerge from trading-up opportunities.  

Next, consider (ii) two rental varieties. From above we immediately get the price for first-time renters in period~$2$, $p_l^2(o) = w/2$. The profit that the supplier obtains from those types is thus equal to the profit that the supplier obtains in (i) from buyer types in the second period. The cut-off type is $w = p_l^1 + \delta p_l^2(l) - \delta p_l^2(o)$,\footnote{Consumer type $v=(v_a,v_b)$ buys in the first period iff $v_i + \delta v_i - p_i^1 - \delta p_i^2(i) \geq \max \{ \delta(v_i - p_i^2(o)), v_j + \delta v_j - p_j^1 - \delta p_j^2(j),\delta(v_j - p_j^2(o)), 0, v_j - p_j^1 + \delta v_i - \delta p_i^2(j), v_i - p_i^1 + \delta v_j - \delta p_j^2(i) \}. $ In the symmetric equilibrium, we are constructing, this condition simplifies to $v_i \geq w$.} where $p_l^2(l)$ is the minimal price in period~$2$, after consumer types have bought the variety with the minimal price in period~$1$. The net present value of the price for repeat renters is $p_l^1 + \delta p_l^2(l)$. The supplier, thus, gets the profit $(1-w)(p_l^1 + \delta p_l^2(l)) + \delta (w-p_l^2)p_l^2$. Plugging in for $w$, we obtain the equivalent problem as in (i): the supplier has to decide when and which types to trade-up. Therefore, $w= (2+\delta)/(4+\delta)$, implying $p_l^2(o)=(2+\delta)/(8+2\delta)$ and $p_l^1 + \delta p_l^2(l) = (2+\delta)^2/(8+2\delta)$. By positive selection, we get $p_l^2(l)=w=(2+\delta)/(4+\delta)$ and hence $p_l^1=p_l^2(o)(2-\delta)=(4-\delta^2)/(8+2\delta)$.\footnote{The undetermined price for switchers or non-switchers $p_i(j)$, respectively $p_i(i)$ is set such that the threshold type is indeed $w$.} Comparing the price levels, we obtain that the seller sets a price in period~$1$ below the static monopoly price, yet if the consumer buys, the price in the second period is above the static monopoly price; if the consumer does not buy, the relevant price falls. I.e. price dynamics emerge.

Finally, consider (iii) mixed varieties. Note that varieties are perfect substitutes. Thus, the supplier can either price the rental variety $a$ as in (ii), and prices for the durable variety $b$ weakly higher as in (i); or, price the durable variety $b$ as in (i), and price the rental variety $a$ weakly higher as in (ii). Either way, the outcome is equivalent as above: pricing dynamics emerge due to trading-up opportunities.

\end{appendix}

%





\end{document}